\documentclass[a4paper,11pt]{article}
\pdfoutput=1

\usepackage{modjheppub} 

\usepackage{xspace}
\usepackage[T1]{fontenc} 

\newcommand{\dtagkx}{D_{\rm tag}K_{\rm frag}X_{\rm frag}}
\newcommand{\ds}{D_{s}}
\newcommand{\dsp}{D_{s}^+}
\newcommand{\dsm}{D_{s}^-}
\newcommand{\dspm}{D_{s}^{\pm}}

\newcommand{\dsst}{D^{\ast}_{s}}
\newcommand{\xfrag}{X_{\rm frag}}

\newcommand{\dtag}{D_{\rm tag}}

\newcommand{\dz}{D^0}
\newcommand{\dc}{D^+}
\newcommand{\lc}{\Lambda_c^+}
\newcommand{\dsmunu}{\ds^+\to\mu^+\nu_{\mu}}

\newcommand{\br}{{\cal B}}
\newcommand{\fds}{f_{\ds}}
\newcommand{\fb}{fb$^{-1}$}

\newcommand{\enu}{e^+\nu_{e}}
\newcommand{\munu}{\mu^+\nu_{\mu}}
\newcommand{\taunu}{\tau^+\nu_{\tau}}
\newcommand{\taumunu}{\tau^+(\mu^+)\nu_{\tau}}
\newcommand{\tauenu}{\tau^+(e^+)\nu_{\tau}}
\newcommand{\taupinu}{\tau^+(\pi^+)\nu_{\tau}}

\newcommand{\mmiss}{M_{\rm miss}}
\newcommand{\fbias}{f_{\rm bias}}
\newcommand{\eecl}{E_{\rm ECL}}

\newcommand{\xdmA}{\xfrag = {\rm nothing}}

\newcommand{\xdmG}{\xfrag = \pi^{\pm}\pi^{\mp}\pi^{0}}

\newcommand{\dskkpi}{\dsp\to K^-K^+\pi^+}
\newcommand{\dskzk}{\dsp\to \overline{K}{}^0K^+}
\newcommand{\kzk}{\overline{K}{}^0K^+}
\newcommand{\dsetapi}{\dsp\to \eta\pi^+}

\newcommand{\gev}{\ensuremath{\mathrm{\,Ge\kern -0.1em V}}\xspace}
\newcommand{\mev}{\ensuremath{\mathrm{\,Me\kern -0.1em V}}\xspace}
\newcommand{\gevc}{\ensuremath{{\mathrm{\,Ge\kern -0.1em V\!/}c}}\xspace}
\newcommand{\mevc}{\ensuremath{{\mathrm{\,Me\kern -0.1em V\!/}c}}\xspace}
\newcommand{\gevcc}{\ensuremath{{\mathrm{\,Ge\kern -0.1em V\!/}c^2}}\xspace}
\newcommand{\gevsqcq}{\ensuremath{{\mathrm{\,Ge\kern -0.1em V^2\!/}c^4}}\xspace}
\newcommand{\mevcc}{\ensuremath{{\mathrm{\,Me\kern -0.1em V\!/}c^2}}\xspace}

\newcommand{\cls}{\ensuremath{\mathrm{CL_s}}\xspace}
\newcommand{\clsb}{\ensuremath{\mathrm{CL_{s+b}}}\xspace}
\newcommand{\clb}{\ensuremath{\mathrm{CL_b}}\xspace}

\def\sPlot{\hbox{$_s$}${\cal P}$lot}

\bellepreprint{Belle Preprint 2013-15}
\kekpreprint{KEK Preprint 2013-25}
\mydate{July 23, 2013}

\title{\boldmath Measurements of branching fractions of leptonic and hadronic $D_s^+$ meson decays and extraction of the $D_s^+$ meson decay constant}

\abstract{
We present measurements of absolute branching fractions of hadronic and leptonic $\ds^+$ decays to $K^-K^+\pi^+$, $\overline{K}{}^0K^+$, 
$\eta\pi^+$, $\mu^+\nu_{\mu}$ and $\tau^+\nu_{\tau}$ and report a search for the leptonic $\dsp\to e^+\nu_{e}$ decays. 
The results are obtained from a data sample of 913~\fb\ collected at or
near the $\Upsilon(4S)$ and $\Upsilon(5S)$ resonances with the Belle detector at the KEKB asymmetric-energy $e^+e^-$ collider. 
The branching fractions of hadronic decays are measured to be
\begin{eqnarray*}
 \br(\ds^+\to K^-K^+\pi^+) & = &(5.06 \pm 0.15\pm 0.21)\%,\\
 \br(\ds^+\to\overline{K}{}^0K^+) & = & (2.95 \pm 0.11\pm 0.09)\%,\\
 \br(\ds^+\to\eta\pi^+) & = & (1.82\pm0.14\pm 0.07)\%,
\end{eqnarray*}
where the first and second uncertainties
are statistical and systematic, respectively. The branching fractions of leptonic decays are measured to be
\begin{eqnarray*}
 \br(\ds^+\to\mu^+\nu_{\mu}) & = &(0.531\pm0.028\pm0.020)\%,\\
 \br(\ds^+\to\tau^+\nu_{\tau}) & = & (5.70\pm0.21{}^{+0.31}_{-0.30})\%,
\end{eqnarray*}
which are combined to determine the $\ds^+$ meson decay constant 
\[
\fds=(255.5\pm4.2\pm5.1)~\mev.  
\]
We find no significant signal for $\ds^+\to e^+\nu_{e}$ decays and set an upper limit of 
$\br(\ds^+\to e^+\nu_{e})<1.0 (0.83)\times 10^{-4}$ at 95\% (90\%) confidence level.
}

\collaboration{The Belle Collaboration}
  \author[23]{A.~Zupanc} 
  \author[13]{I.~Adachi} 
  \author[58]{H.~Aihara} 
  \author[3]{K.~Arinstein} 
  \author[45]{D.~M.~Asner} 
  \author[20]{T.~Aushev} 
  \author[52]{A.~M.~Bakich} 
  \author[46]{A.~Bala} 
  \author[15]{B.~Bhuyan} 
  \author[64]{G.~Bonvicini} 
% ...

% ...
\affiliation[3]{Budker Institute of Nuclear Physics SB RAS and Novosibirsk State University, Novosibirsk 630090, Russian Federation}
% ...
\affiliation[13]{High Energy Accelerator Research Organization (KEK), Tsukuba 305-0801, Japan}
% ...
\affiliation[15]{Indian Institute of Technology Guwahati, Assam 781039, India}
% ...
\affiliation[20]{Institute for Theoretical and Experimental Physics, Moscow 117218, Russian Federation}
% ...
\affiliation[23]{Institut f\"ur Experimentelle Kernphysik, Karlsruher Institut f\"ur Technologie, 76131 Karlsruhe, Germany}
% ...
\affiliation[45]{Pacific Northwest National Laboratory, Richland, WA 99352, USA}
\affiliation[46]{Panjab University, Chandigarh 160014, India}
% ...
\affiliation[52]{School of Physics, University of Sydney, NSW 2006, Australia}
% ...
\affiliation[58]{Department of Physics, University of Tokyo, Tokyo 113-0033, Japan}
% ...
\affiliation[64]{Wayne State University, Detroit, MI 48202, USA}
% ...

\begin{document} 
\maketitle
\flushbottom

\section{Introduction}
Precise determination of the Cabibbo-Kobayashi-Maskawa (CKM) quark mixing matrix 
leads to a deeper understanding of the flavor structure in the 
Standard Model (SM) and provides a portal to New Physics (NP) processes at higher energy scales. Many of the constraints
on the CKM unitarity triangle given by the precise experimental results on decays of $B$ mesons (see ref.~\cite{Brodzicka:2012jm} 
for a review of results from the Belle collaboration) rely on lattice gauge theory (LQCD) calculations of quantities that parameterize 
nonperturbative QCD contributions to weak decays and mixing (see section 17 in ref.~\cite{Beringer:1900zz} for a review). 
Among these quantities, the pseudoscalar meson decay constants play an important role --- without them, for example, 
an interpretation of measurements of purely leptonic decays $B^+\to \tau^+\nu_{\tau}$ \cite{Adachi:2012mm,Hara:2010dk} 
and $B_s^0\to \mu^+\mu^-$ \cite{Aaij:2012nna} that are particularly sensitive to NP contributions is not possible. 
In some NP scenarios, the leptonic decay rates of the $\dsp$ mesons could also be modified although the expected effects are smaller 
than in the $B$ meson sector \cite{Akeroyd:2007eh, Akeroyd:2009tn, Barranco:2013tba, Crivellin:2013wna}. 
Measurements of leptonic decays of charmed hadrons, $D_{s}^+\to\ell^+\nu_{\ell}$ where $\ell^+=e^+$, $\mu^+$ or $\tau^+$, 
therefore enable precision tests of LQCD calculations of decay constants performed in the charm sector and can provide 
additional constraints on NP.\footnote{Charge conjugation is assumed throughout this paper unless stated otherwise.}

Purely leptonic decays of mesons are among the simplest and theoretically cleanest processes. The branching fraction of
$\dsp$ meson leptonic decays that proceed via the mutual annihilation of the $c$ and $\overline{s}$-quarks into a 
virtual $W^+$ boson is given in the SM by 
\begin{equation}
 \br(D_{s}^+\to \ell^+\nu_{\ell})=\frac{\tau_{\ds}m_{D_{s}}}{8\pi}f_{D_{s}}^2G_F^2|V_{cs}|^2m_{\ell}^2\left(1-\frac{m_{\ell}^2}{m_{D_{s}}^2} \right)^2.
 \label{eq:brleptonic_sm}
\end{equation}
Here, $m_{D_{s}}$ is the $D_{s}^+$ meson mass, $\tau_{\ds}$ is its lifetime, $m_{\ell}$ is the lepton mass, $V_{cs}$ is the relevant
CKM matrix element, and $G_F$ is the Fermi coupling constant. The parameter $f_{D_{s}}$ is the $\dsp$ meson decay constant and is 
related to the wave-function overlap of the meson's constituent quark and anti-quark. The leptonic decays of pseudoscalar mesons 
are suppressed by helicity conservation and their decay rates are thus proportional to the square of the charged lepton mass. 
Leptonic $\dsp$ decays into electrons with $\br\sim 10^{-7}$ are not observable yet whereas decays to taus are favored over 
decays to muons. In particular, the ratio of the latter decays is equal to
$R^{\ds}_{\tau/\mu}\equiv \br(\dsp\to\tau^+\nu_{\tau})/\br(\dsp\to\mu^+\nu_{\mu})=m_{\tau}^2/m_{\mu}^2\cdot(1-m^2_{\tau}/m^2_{\ds})^2/(1-m^2_{\mu}/m^2_{\ds})^2=9.762\pm0.031$, 
based on the world average masses of the muon, tau and $\dsp$ meson given in ref. \cite{Beringer:1900zz}. 
Any deviation from this expectation could only be interpreted as violation of lepton universality in charged 
currents and would hence point to NP effects \cite{Filipuzzi:2012mg}.

In the context of the SM, a measurement of $\br(D_{s}^+\to \ell^+\nu_{\ell})$ determines the $\dsp$ meson decay constant 
since the magnitude of the CKM matrix element $|V_{cs}|$ is precisely determined from other measurements and the assumption 
that the CKM matrix is unitary. Measurements of $\fds$ have been made previously by several groups: 
CLEO \cite{Alexander:2009ux,Naik:2009tk,Onyisi:2009th}, Belle \cite{Widhalm:2007ws} and BaBar \cite{delAmoSanchez:2010jg}. 
The current world average is $\fds^{\rm exp}=(260.0\pm5.4)$~\mev~\cite{Beringer:1900zz}. 
Within the SM, $\fds$ has been predicted using several 
methods \cite{Davies:2010ip,Bazavov:2011aa,Becirevic:2013mp,Blossier:2009bx,Bordes:2005wi,Lucha:2011zp,Badalian:2007km,Hwang:2009qz} 
and most calculations give values lower 
than the $\fds$ measurement although within theoretical and experimental uncertainties. The largest discrepancy is with an 
unquenched LQCD calculation that yields $\fds^{\rm LQCD}=(248.0\pm2.5)$~\mev \cite{Davies:2010ip}. Measurements of $\fds$ with an accuracy 
that matches the precision of theoretical calculations are thus necessary to check and further constrain theoretical methods.

Hadronic decays, $\dskkpi$ and $\dskzk$, are the reference modes for the measurements of branching fractions of the $\dsp$ decays to
any other final state \cite{Beringer:1900zz}. In addition, precise measurements of the absolute hadronic $\dsp$ meson branching fractions
improve our knowledge of the $B_{(s)}$ decays involving $\dsp$, such as $B_{(s)}^0\to D_{s}^{(\ast)-}D_{(s)}^{(\ast)+}$ \cite{Beringer:1900zz}, 
and of most of the other branching fraction measurements of $B_{s}$ mesons performed at LHCb, like 
$B^0_s\to \mu^+\mu^-$ \cite{Aaij:2012nna}. For $B_s$-decay branching fraction measurements performed at LHCb, 
the key systematic uncertainty \cite{Fleischer:2010ay} is the ratio of fragmentation fractions $f_s/f_d$, whose experimental
systematic error is dominated by $\br(\dskkpi)$ \cite{Aaij:2011jp,Aaij:2013qqa}.\footnote{The fragmentation fractions, $f_q$,
describe the probability that a $b$ quark fragments in a $B_{q}$ meson, where $q=d$ or $s$.} Normalization branching fractions, 
$\br(\dskkpi)$ and $\br(\dskzk)$, have been measured so far only by CLEO \cite{Alexander:2008aa} (see also the very recent update 
in ref.~\cite{Onyisi:2013pua}). It is therefore important to provide new and independent measurements.

In this paper, we present results of absolute branching fraction measurements of $\ds^+\to\mu^+\nu_{\mu}$ and $\ds^+\to\tau^+\nu_{\tau}$ 
decays and perform a search for $\ds^+\to e^+\nu_{e}$ decays. The measurement of $\br(\ds^+\to\mu^+\nu_{\mu})$ presented here supersedes 
the previous Belle measurement~\cite{Widhalm:2007ws}.
The analysis described here has a number of significant improvements, including an increased data sample and significantly improved 
inclusive $\dsp$ reconstruction efficiency. The combined effect of these improvements and the accompanying change in the extraction of 
relevant signal yields results in a reduction of the expected error of $\br(\ds^+\to\mu^+\nu_{\mu})$ by more than a factor of two. The 
new analysis has improved systematic uncertainties. In addition, we present first measurements of absolute branching fractions of 
the $\dsp$ normalization decays, $\dskkpi$ and $\dskzk$, and of $\dsetapi$ decays. This analysis is based on a data sample of $913$~fb$^{-1}$ 
recorded at and near the $\Upsilon(4S)$ and $\Upsilon(5S)$ resonances ---  well above the open charm threshold --- by the Belle detector 
at the KEKB asymmetric-energy collider \cite{Kurokawa:2001nw,AbeKEKB}. 

The rest of the paper is structured as follows. We describe the Belle detector and the data sample in 
section \ref{sec:belle:data}. In section \ref{sec:method}, we present the method of measuring the absolute branching
fraction of $\dsp$ decays. The inclusive and exclusive event reconstruction steps are described in sections 
\ref{sec:dsinclusive} and \ref{sec:exclusiverec}, respectively. Determination of the absolute branching fractions is discussed 
in section \ref{sec:absbr}. Systematic uncertainties are itemized in section \ref{sec:syst}. We summarize our results in 
section \ref{sec:results} and conclude in section \ref{sec:conclusion}.

\section{Belle detector and data sample}
\label{sec:belle:data}

The data used in this analysis were collected with the Belle detector at the KEKB asymmetric energy $e^+e^−$ collider. The 
Belle detector is a large-solid-angle magnetic spectrometer that consists of a silicon vertex detector (SVD), a 50-layer 
central drift chamber (CDC), an array of aerogel threshold Cherenkov counters (ACC), a barrel-like arrangement of time-of-flight 
scintillation counters (TOF), and an electromagnetic calorimeter (ECL) comprised of CsI(Tl) crystals located inside a superconducting 
solenoid coil that provides a 1.5~T magnetic field.  An iron flux-return located outside of the coil is instrumented to detect 
$K_L^0$ mesons and to identify muons (KLM). The detector is described in detail elsewhere~\cite{Brodzicka:2012jm,Abashian:2000cg}. 
Two inner detector configurations were used. A 2.0 cm diameter beampipe and a 3-layer silicon vertex detector was used for the first 
sample of 156 fb$^{-1}$, while a 1.5 cm diameter beampipe, a 4-layer silicon detector and a small-cell inner drift chamber were used 
to record the remaining 757 fb$^{-1}$.

Charged particles are reconstructed with the CDC and the SVD. Each is required to have an impact parameter with 
respect to the interaction point (IP) of less than 1.5~cm along the beam direction ($z$) and less than 0.5~cm in 
the transverse ($r-\phi$) plane. 
A likelihood ratio for a given track to be a kaon or pion, ${\cal L}_{(K, \pi)}$, is obtained by utilizing 
specific ionization energy loss measurements in the CDC, light yield measurements from the ACC, and time-of-flight 
information from the TOF. For electron identification, we use position, cluster energy, and shower
shape in the ECL, combined with track momentum and $dE/dx$ measurements in the CDC and hits in the ACC.
For muon identification, we extrapolate the CDC track to the KLM and compare the measured range and transverse 
deviation in the KLM with the expected values. Photons are detected with the ECL and are required to have energies in the laboratory
frame of at least 50 (100) \mev in the ECL barrel (endcaps). Neutral pion candidates are 
reconstructed using photon pairs with an invariant mass between 120 and 150~\mevcc, which corresponds to  
$\pm$3.2~$\sigma$ around the nominal $\pi^0$ mass~\cite{Beringer:1900zz}, where $\sigma$ represents 
the invariant mass resolution. Neutral kaon candidates are reconstructed using charged pion pairs 
with an invariant mass within $\pm20$~\mevcc ($\pm5~\sigma$) of the nominal $K^0$ mass.

We use Monte Carlo (MC) events generated with EVTGEN~\cite{Lange:2001uf} and JETSET~\cite{Sjostrand:1993yb} and then processed through 
the detailed detector simulation implemented in GEANT3~\cite{Brun:1987ma}. QED final state radiation from charged particles
is added during generation using the PHOTOS package~\cite{Barberio:1993qi}. The simulated samples for $e^+e^-$ annihilation 
to $q\overline{q}$ ($q=u$, $d$, $s$, $c$, and $b$) are equivalent to six times the integrated luminosity of the data and are used to develop
methods to separate signal events from backgrounds, identify types of background events, determine reconstruction efficiencies
and the distributions needed for the extraction of the signal decays.

\section{Method overview}
\label{sec:method}
The method of absolute branching fraction measurement of $\dsm$ decays is similar to the one previously used
by Belle \cite{Widhalm:2006wz,Widhalm:2007ws} and BaBar \cite{delAmoSanchez:2010jg}.
In this method, the $e^+e^- \to c\bar{c}$ events that contain $\ds^-$ mesons produced through the reactions
\begin{equation}
 e^+e^-\to c\bar{c}\to \dtagkx\ds^{\ast -},\qquad\ds^{\ast -}\to\ds^-\gamma, 
 \label{eq:signal_events_type}
\end{equation}
are fully reconstructed in two steps. In these events, one of the two charm quarks hadronizes into a $\ds^{\ast -}$ meson while the other
hadronizes into a tagging charm hadron, denoted $\dtag$, that is one of $\dz$, $\dc$, $\lc$, $D^{\ast +}$ or $D^{\ast 0}$.  
The strangeness of the event is conserved by requiring an additional kaon, denoted $K_{\rm frag}$, be produced in the fragmentation process;
$K_{\rm frag}$ is either $K^+$ or $K^0_S$. In events where $\lc$ is the tagging charm hadron, 
the baryon number of the event is conserved by requiring an anti-proton. Since Belle collected data at energies well above the
$D{}^{(\ast)}_{\rm tag} K_{\rm frag} D_s^{\ast-}$ threshold, additional particles can be produced in the course of hadronization. 
These particles are denoted as $\xfrag$ and consist of an even number of kaons plus any number of pions or photons. In this measurement,
only pions are considered when reconstructing the fragmentation system $\xfrag$.\footnote{The strangeness-conserving kaon and 
the baryon-number-conserving anti-proton are counted separately and are not included in the $\xfrag$ system.} We require $\ds^-$ mesons to be produced in a
$\ds^{\ast -}\to\ds^-\gamma$ decay, which provides a powerful kinematic constraint that improves the resolution of the missing mass (defined below)
and suppresses the combinatorial background.

In the first step of the measurement, no requirements are placed on the daughters of the signal $\ds^-$ meson
in order to obtain an inclusive sample of $\ds^-$ events that is used for normalization
in the calculation of the branching fractions. The number of inclusively reconstructed $\dsm$ mesons is
extracted from the distribution of events in the missing mass, $\mmiss(\dtagkx\gamma)$, recoiling against the $\dtagkx\gamma$ system
\begin{equation}
\mmiss(\dtagkx\gamma)  =  \sqrt{p_{\rm miss}(\dtagkx\gamma)^2},
\label{eq:massds} 
\end{equation}
where $p_{\rm miss}$ is the missing four-momentum in the event
\begin{equation}
p_{\rm miss}(\dtagkx\gamma)  =  p_{e^+} + p_{e^-} - p_{\dtag} - p_{K_{\rm frag}} - p_{\xfrag} - p_{\gamma}.\\
\label{eq:pmiss}
\end{equation}
Here, $p_{e^+}$ and $p_{e^-}$ are the known four-momenta of the colliding positron and electron beams, respectively, and $p_{\dtag}$, 
$p_{K_{\rm frag}}$, $p_{\xfrag}$, and $p_{\gamma}$ are the measured four-momenta of the reconstructed $\dtag$, 
strangeness-conserving kaon, fragmentation system and the photon from
$\ds^{\ast-}\to\dsm\gamma$, respectively. Correctly reconstructed events described in eq.~(\ref{eq:signal_events_type}) produce a peak in the 
$\mmiss(\dtagkx\gamma)$ at the nominal $\dsm$ meson mass.

In the second step of the analysis, we search for the decay products of a specific $\dsm$ meson decay within the inclusive $\dsm$ 
meson sample reconstructed in the first step. In particular, we reconstruct purely leptonic 
$\dsm\to e^-\overline{\nu}{}_{e}$, $\dsm\to\mu^-\overline{\nu}{}_{\mu}$, and $\dsm\to\tau^-\overline{\nu}_{\tau}$ decays within the 
inclusive $\dsm$ sample by requiring an additional charged track identified as an electron, muon or charged pion
in the rest of the event. In the case of $\dsm\to\tau^-\overline{\nu}{}_{\tau}$ decays, the electron, muon or pion 
identifies the subsequent tau decay to $e^-\overline{\nu}{}_e\nu_{\tau}$, $\mu^-\overline{\nu}_{\mu}\nu_{\tau}$ or $\pi^-\nu_{\tau}$,
respectively. Hadronic decays, $\dsm\to K^0K^-$ and $\eta\pi^-$, are reconstructed partially by explicitly requiring only the charged kaon 
or pion (originating directly from $\dsm$ meson decay) in the rest of the event but with no requirements on the 
neutral hadrons ($K^0$ or $\eta$) in order to increase the reconstruction efficiency. In the case of $\dsm\to K^-K^+\pi^-$, all three 
charged tracks are required in the rest of the event. More details are given in section \ref{sec:exclusiverec}.

\section{\boldmath Inclusive $\dspm$ reconstruction}
\label{sec:dsinclusive}
The reconstruction of the inclusive $\dspm$ sample starts with the reconstruction of the tagging charmed hadron, $\dtag$. 
To maximize the reconstruction efficiency with reasonable purity, the ground-state $\dtag$ hadrons ($\dz$, $\dc$, $\lc$) 
are reconstructed in the 18 hadronic decay modes listed in table \ref{tab:dtag_modes}. Only modes with up to one $\pi^0$ 
in the final state are used to avoid large backgrounds. If $\dtag$ is reconstructed as $\lc$ baryon, an additional anti-proton
is required in order to conserve the baryon number in the event.
\begin{table}[t]
\centering
  \begin{tabular}{l|c}
  $\dz$ modes			& ${\cal B}~[\%]$\\\hline
  $K^-\pi^+$			& 3.9 \\
  $K^-\pi^+\pi^0$		& 13.9 \\
  $K^-\pi^+\pi^+\pi^-$		& 8.1 \\
  $K^-\pi^+\pi^+\pi^-\pi^0$	& 4.2 \\
  $K^0_S\pi^+\pi^-$		& 2.9 \\
  $K^0_S\pi^+\pi^-\pi^0$	& 5.4 \\ \hline
  Sum                           & 38.4
  \end{tabular}  
\hspace{0.05\textwidth}
    \begin{tabular}{l|c}
   $\dc$ modes			& ${\cal B}~[\%]$\\\hline
   $K^-\pi^+\pi^+$		& 9.4 \\
   $K^-\pi^+\pi^+\pi^0$		& 6.1 \\
   $K^0_S\pi^+$			& 1.5 \\
   $K^0_S\pi^+\pi^0$		& 6.9 \\
   $K^0_S\pi^+\pi^+\pi^-$	& 3.1 \\
   $K^+K^-\pi^+$		& 1.0 \\ \hline
   Sum                          & 28.0
  \end{tabular}
\hspace{0.05\textwidth}
    \begin{tabular}{l|c}
   $\lc$ modes			& ${\cal B}~[\%]$\\\hline
   $pK^-\pi^+$			& 5.0 \\
   $pK^-\pi^+\pi^0$		& 3.4 \\
   $pK^0_S$			& 1.1 \\
   $\Lambda\pi^+$			& 1.1 \\
   $\Lambda\pi^+\pi^0$		& 3.6 \\
   $\Lambda\pi^+\pi^+\pi^-$	& 2.6 \\ \hline
   Sum                          & 16.8
  \end{tabular}
\caption{Summary of $\dtag=\dz$, $\dc$ and $\lc$ decay modes used in this measurement. The branching fractions are taken
from ref.~\cite{Beringer:1900zz}.}
\label{tab:dtag_modes}
 \end{table}

The magnitude of the center-of-mass (CMS) momentum of the $\dtag$ candidates is required to be greater than 2.3~\gevc (or 2.5~\gevc for the 
less clean $\dtag$ modes) to reduce the combinatorial background and $e^+e^-\to B\overline{B}$ events. The decay products of the 
$\dtag$ candidate are fitted to a common vertex; candidates with a poor fit quality are discarded by requiring 
$\chi^2/{\rm n.d.f.}<20$, where n.d.f. is the number of degrees of freedom of the kinematic fit. The purity of the $\dtag$ sample,
given as a fraction of correctly reconstructed $\dtag$ candidates, is rather low at this stage --- around 17\% in the signal region,
defined as $\pm3~\sigma$ interval around the nominal $\dtag$ mass, where $\sigma$ is the $\dtag$ decay-mode-dependent invariant mass 
resolution that ranges from 4 to 12~\mevcc. To further purify the $\dtag$ sample, we train a NeuroBayes~\cite{Feindt:2006pm} neural 
network using a small sample of data (0.7\% of the total sample). The network combines information from the following input variables
into a single variable: the distance between the decay and the production vertices of the $\dtag$ candidate in the $r-\phi$ plane, where
the $\dtag$ production vertex is defined by the intersection of its trajectory with the IP region; the $\chi^2/{\rm n.d.f.}$ of 
the vertex fit of the $\dtag$ candidate; the cosine of the angle between the $\dtag$ momentum and the vector joining
its decay and production vertices in the $r-\phi$ plane; for two-body decays, the cosine of the angle between the momentum 
of either $\dtag$ daughter and the boost direction of the laboratory frame in the $\dtag$ rest frame; the particle identification 
likelihood ratios; and, for the $\dtag$ decay modes with a $\pi^0$, the smaller of the two photon energies. To obtain the signal 
and background distributions of the network's input variables, a statistical tool to unfold the data distributions 
(\sPlot~\cite{Pivk:2004ty}) is applied. Network is then applied to the complementary subsample (again representing around 
0.7\% of the total sample) that we use to optimize the selection on the network output variable for each $\dtag$ mode 
individually by maximizing $S/\sqrt{S+B}$, where $S$ ($B$) refers to the signal (background) yield in the signal 
window of $\dtag$ invariant mass determined by performing a fit to the $\dtag$ invariant mass 
distribution.\footnote{
This approach avoids a bias of the selection originating from statistical fluctuations possibly learned by the network. Since
the optimization of $\dtag$'s selection is performed using a very small fraction of data, any bias that could be triggered by 
statistical fluctuations is negligible.} 
After the optimization, the purity of the correctly reconstructed $\dtag$ candidates increases from 17\% to 42\% while 
only 16\% of signal $\dtag$ candidates is lost. We retain only $\dtag$ candidates from the signal region of the $\dtag$ 
invariant mass in the rest of the analysis.

Once the ground-state $\dtag$ hadrons have been reconstructed, $\dz$ and $\dc$ mesons originating from
$D^{\ast}$ decays are identified by reconstructing the decays $D^{\ast +}\to \dz\pi^+$, $\dc\pi^0$, and 
$D^{\ast 0}\to \dz\pi^0$, $\dz\gamma$. We do this to purify the subsequent 
$K_{\rm frag}\xfrag\gamma$ reconstruction: by absorbing one more particle into the tagging charm hadron, the subsequent 
combinatorial background is reduced. In addition, by reconstructing $D^{\ast +}\to \dz\pi^+$ decays, we 
can determine the flavor or charm quantum number of $\dz$ or $\overline{D}{}^{0}$ candidates reconstructed in final states with a $K^0_S$. 
The pion from 
the $D^{\ast}$ decay is refitted to the $D$ production vertex to improve the resolution of the mass 
difference, $\Delta M = M(D\pi)-M(D)$. The laboratory frame energy of the photon(s) originating from the $\pi^0$  
produced in $D^{\ast}\to D\pi^0$ (produced directly in $D^{\ast 0}\to \dz\gamma$) is required to be larger than 50 (175)~\mev. 
In the $\dz\gamma$ final state, the $\gamma$ candidate is combined with each 
other photon and, if the two-photon invariant mass is within 10~\mevcc around the nominal $\pi^0$ mass and their energy asymmetry 
($(E_{\gamma_1}-E_{\gamma_2})/(E_{\gamma_1}+E_{\gamma_2})$) is smaller than 0.5, the $D^{\ast 0}$ candidate is 
rejected. For all $D^{\ast}$ decays, the mass difference, $\Delta M$, is required to be within 3~$\sigma$ 
of the corresponding nominal mass difference.

For strangeness conserving kaon candidate, $K_{\rm frag}$, all $K^{\pm}$ or $K^0_S$ candidates 
that do not overlap with the $\dtag$ candidate are considered. 

From the remaining tracks and $\pi^0$ candidates in the event that do not overlap with the $\dtag K_{\rm frag}$ candidate, we form the 
$\xfrag$ candidates. Only modes with up to three pions and up to one $\pi^0$ are used to suppress the combinatorial background. 
In addition, pions must have a momentum above 100~\mevc in the laboratory frame. At this stage, no requirement is applied to the 
total charge of the $\xfrag$ system.

The $\dtag$, $\xfrag$ and $K_{\rm frag}$ candidates are combined to form $\dtagkx$ combinations. 
We keep only combinations with total charge $\pm 1$; these constitute the inclusive sample
of $\ds^{\ast\mp}$ mesons. The charm and strange quark content of the 
$\dtagkx$ system is required to be consistent with that recoiling from a $\dsst$: if $\dtag$ is reconstructed in 
a flavor-specific decay mode and $K_{\rm frag}$ is charged, the kaon
charge and the charm quantum number of $\dtag$ must be opposite the $\dsst$ charge; if $K_{\rm frag}$ is neutral 
the charm quantum number of $\dtag$ must be opposite the $\dsst$ charge; and if $\dtag$ is reconstructed in a 
self-conjugate decay mode, the charge of $K_{\rm frag}$ must be opposite the $\dsst$ charge. All other candidates are rejected.
A kinematic fit to $\dtagkx$ candidates is performed in which the particles are required to originate from a common
point within the IP region and the $\dtag$ mass is constrained to its nominal value. We select only one $\dtagkx$ candidate per 
event that has its missing mass, $\mmiss(\dtagkx)=\sqrt{|p_{e^+} + p_{e^-} - p_{\dtag} - p_{K_{\rm frag}} - p_{\xfrag}|^2}$, 
closest to the nominal $\ds^{\ast+}$ mass and between $2.00$ and $2.25$~\gevcc, which corresponds to a $\pm3~\sigma$ interval. 

Finally, a photon candidate is identified that is consistent with the decay $\ds^{\ast \pm}\to \dspm\gamma$ and
does not overlap with the $\dtagkx$ system. We require that the energy of the photon candidate be larger than 120~\mev in the 
laboratory frame and that the cosine of the angle between the CMS momenta of the $\dtag$ hadron and the photon candidate be negative, 
since the signal photon should be in the hemisphere opposite the $\dtag$ hadron. We perform a similar kinematic fit
with the signal photon included and with the missing mass recoiling against the $\dtagkx$ system constrained to the nominal 
$\ds^{\ast+}$ mass. All $\dtagkx\gamma$ candidates are required to have a CMS momentum larger than 2.8~\gevc 
and $\mmiss(\dtagkx\gamma)>1.83$~\gevcc (see eq.~(\ref{eq:massds})). After the final selections, there are an average of 
2.1 $\dtagkx\gamma$ candidates per event; these are due solely to multiple $\gamma$ candidates. Among these, we select 
the one with the highest NeuroBayes network output that is trained to separate signal photons from others based on
photon energy, the detecting region of the ECL (forward, barrel or backward region), the ratio of the energies summed in 
$3\times3$ and $5\times5$ ECL crystals in the transverse plane around the crystal with the largest energy deposit, 
the invariant mass of the combination of the photon candidate with any other photon candidate that is closest to the 
$\pi^0$ nominal mass, the energy asymmetry of this two-photon combination, and the invariant mass and energy asymmetry of the 
two-photon combination whose invariant mass is second closest to the nominal $\pi^0$ mass. 
A relative gain of 23\% in absolute reconstruction efficiency is obtained by applying the best $\dtagkx\gamma$ candidate 
selection instead of a completely random selection. Figure \ref{fig:dsincl:datafitres} shows the distributions of 
$\mmiss(\dtagkx\gamma)$ for each $\xfrag$ mode. 

\subsection{\bf\boldmath Inclusive $\dsp$ yield extraction}
\label{sec:inclds:yield:extraction}
The yield of inclusively reconstructed $\dsp$ mesons is determined by performing a $\chi^2$ fit to the missing mass $\mmiss(\dtagkx\gamma)$ 
distribution for each $\xfrag$ mode. The events fall into six categories: signal candidates; 
mis-reconstructed signal candidates, where either $K_{\rm frag}$ or one of the pions forming the $\xfrag$ system originates in 
reality from a $\dsp$ decay; background candidates where the signal $\gamma$ candidate originates from $D^{\ast0}\to D^0\gamma$ decays; 
background candidates where the signal $\gamma$ originates from the $\pi^0$ produced in $D_{(s)}^{\ast}\to D_{(s)}\pi^0$ decays; 
background candidates with a bad $\gamma$ -- the energy deposited in the ECL being produced by an unmatched charged track or by a 
beam-induced interaction; and background candidates where the signal $\gamma$ originates from a $\pi^0$ that does not itself originate 
from a $D^{\ast}_{(s)}$ decay. Each of the six categories is represented with a 
smoothed non-parametric histogram~\cite{Blobel} probability density function (PDF), 
${\cal H}(\mmiss(\dtagkx\gamma))$, taken from a large sample of MC events. The fit function for a given $\xfrag$ mode is written as
\begin{eqnarray}	
 {\cal F}^{\xfrag}(\mmiss(\dtagkx\gamma)) & = & N^{\xfrag}_{\rm sig}{\cal H}_{\rm sig}^{\xfrag}(\mmiss(\dtagkx\gamma)-\delta_{\mmiss})\otimes {\cal G}(\sigma_{\rm cal}) \nonumber\\
 & & + \sum_{i=1}^5 N_{i}^{\xfrag}{\cal H}_{i}^{\xfrag}(\mmiss(\dtagkx\gamma)),
 \label{eq:mmiss:fitfdata}
\end{eqnarray}
where $N$ represents the yield of each component and the first (second) term describes 
the contribution of signal (the sum of the five background components). The histogram PDF of the signal, 
${\cal H}_{\rm sig}^{\xfrag}(\mmiss(\dtagkx\gamma)-\delta_{\mmiss})$, is numerically 
convolved with a Gaussian function, ${\cal G}(\sigma_{\rm cal})$, centered at zero and with width $\sigma_{\rm cal}$, 
which takes into account possible differences between $\mmiss(\dtagkx\gamma)$ resolutions in the data and MC 
samples. The calibration of $\sigma_{\rm cal}$ is described in the next paragraph. 
The position of the signal peak in data relative to the position in the MC, $\delta_{\mmiss}$, is a free parameter of the fit. We 
also float all normalization parameters, $N^{\xfrag}_{i}$, except the normalization of the background component where 
the signal $\gamma$ candidate originates from $D^{\ast0}\to D^0\gamma$ decays, which is fixed relative to the more abundant and
similar background component where the signal $\gamma$ candidate originates from the $\pi^0$ produced in 
$D^{\ast0}_{(s)}\to D_{(s)}\pi^0$ decays. The fraction $f_{D^0\gamma/D_{(s)}\pi^0}$ is fixed to the value obtained in the MC sample.

We calibrate the $\mmiss(\dtagkx\gamma)$ resolution using the mass difference between $\ds^{\ast+}$ and $\dsp$, 
$\Delta M = M_{\ds^{\ast+}} - M_{\dsp}$, for exclusively reconstructed $\ds^{\ast+}\to\dsp\gamma$ decays, where $\dsp$ decays 
to $\phi\pi^+$ and $\phi\to K^+K^-$. In the exclusive reconstruction of $\ds^{\ast+}$ mesons, the same requirements 
are used for the signal photon candidate as in the inclusive reconstruction. The dominant contribution to 
the $\Delta M$ and $\mmiss(\dtagkx\gamma)$ resolutions is the signal photon energy resolution. In the former 
case, the smearing of the $\dsp$ momentum cancels almost completely in the mass difference while, in the latter 
case, the impact of experimental smearing of $p_{\rm miss}(\dtagkx)$ on $\mmiss(\dtagkx\gamma)$ is minimized 
by performing a mass constrained vertex fit of $\dtagkx$ candidates to the nominal $\ds^{\ast+}$ mass. According to 
the MC study, the $\Delta M$ and $\mmiss(\dtagkx\gamma)$ resolutions are the same to within a few percent, which 
justifies the calibration of the $\mmiss(\dtagkx\gamma)$ resolution by comparing $\Delta M$ resolutions of exclusively 
reconstructed $\ds^{\ast+}\to\dsp\gamma$ decays obtained from data and MC. We parameterize the contribution of correctly
reconstructed $\ds^{\ast+}\to\dsp\gamma$ decays in $\Delta M$ as in the case of the signal parameterization of $\mmiss(\dtagkx\gamma)$
as a histogram PDF convolved with a Gaussian function, ${\cal H}_{\rm sig}(\Delta M - \delta_{\Delta M})\otimes {\cal G}(\sigma_{\rm cal})$,
where $\delta_{\Delta M}$ is the difference between the peak positions in data and MC.
The background shape is fitted by a second order polynomial. We achieve best agreement between data and 
simulated $\Delta M$ distributions when $\sigma_{\rm cal}=2.0\pm0.2$~\mevcc. 

\begin{figure}[hbt!]
 \centering
 \includegraphics[width=0.475\textwidth]{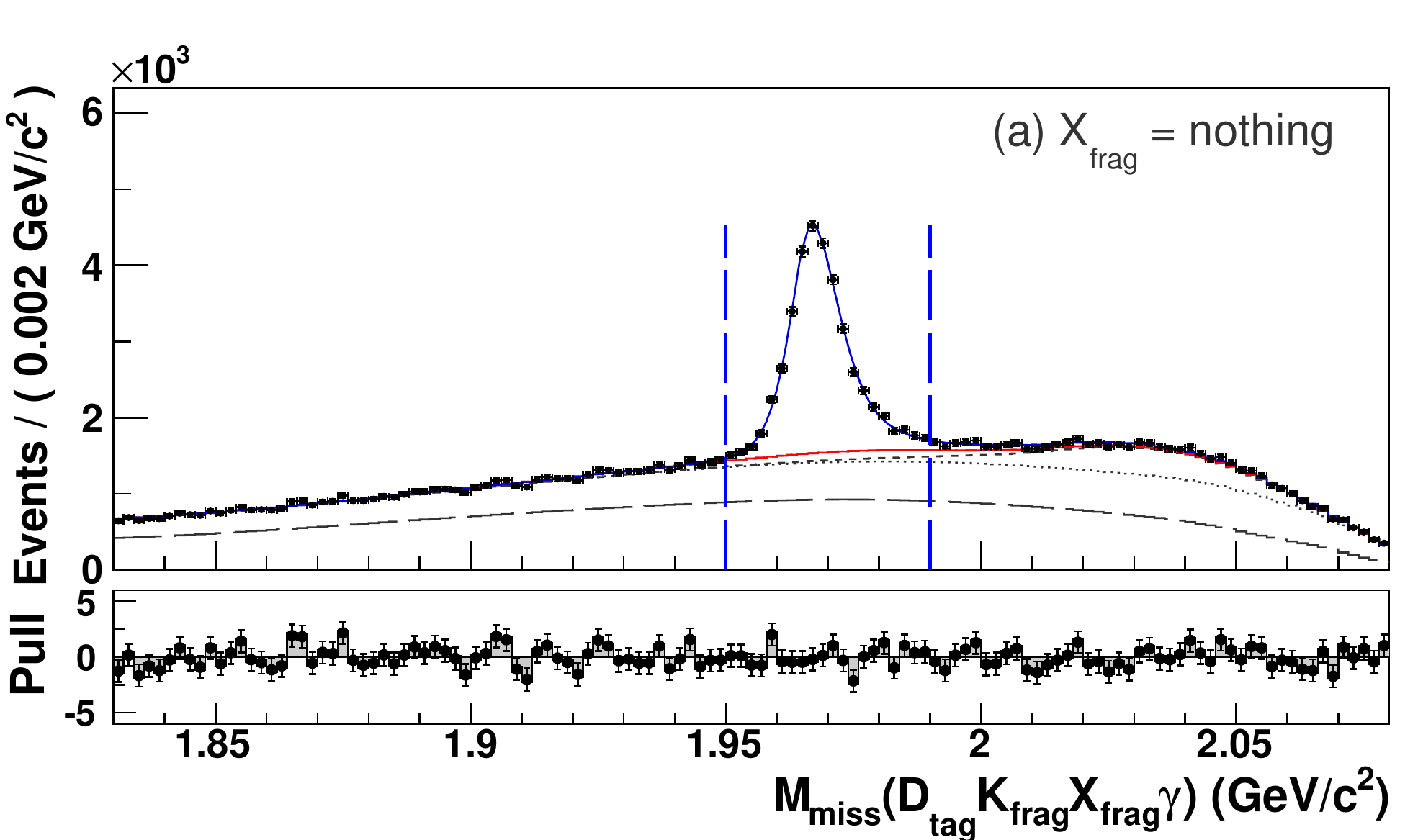}\\
 \includegraphics[width=0.475\textwidth]{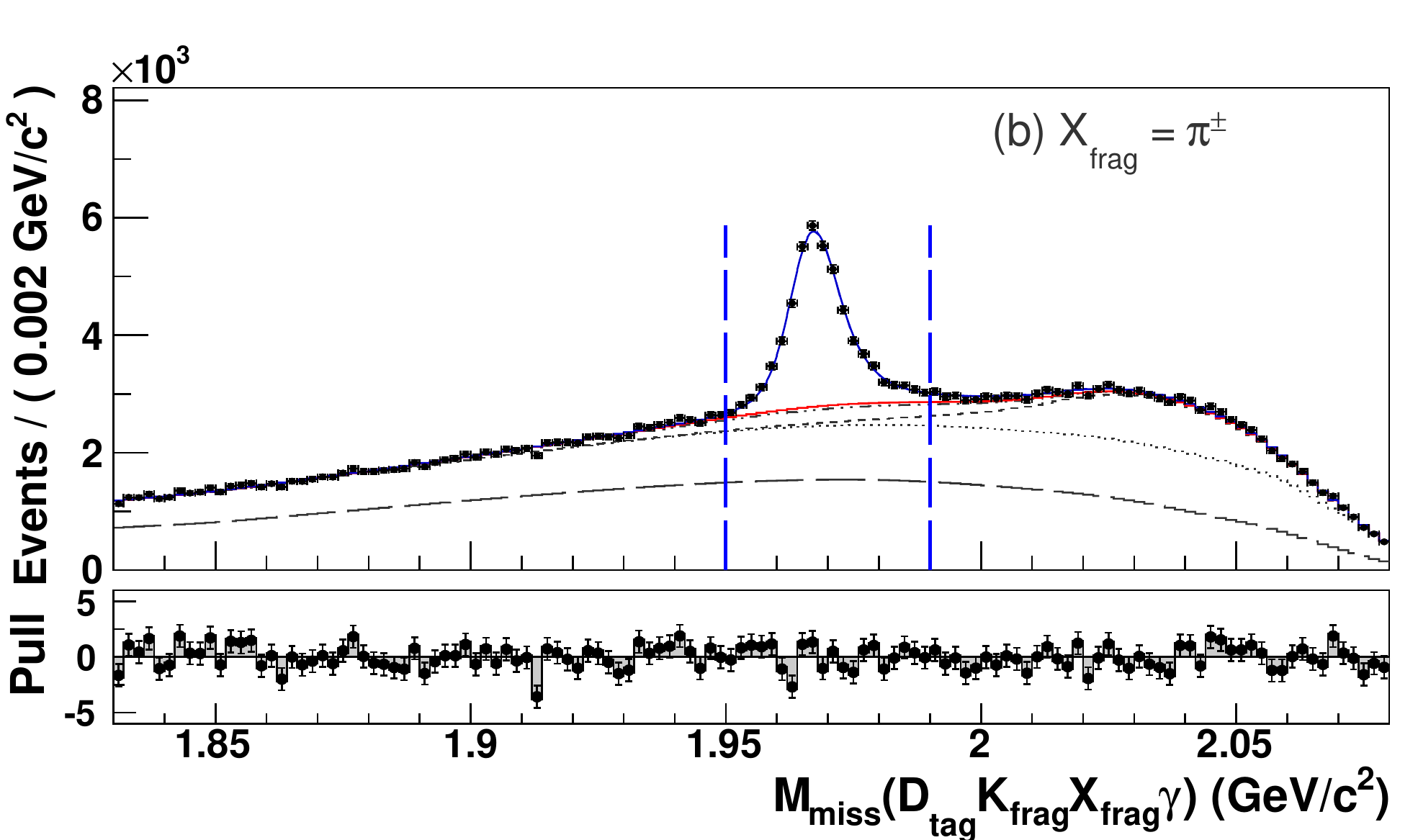}
 \includegraphics[width=0.475\textwidth]{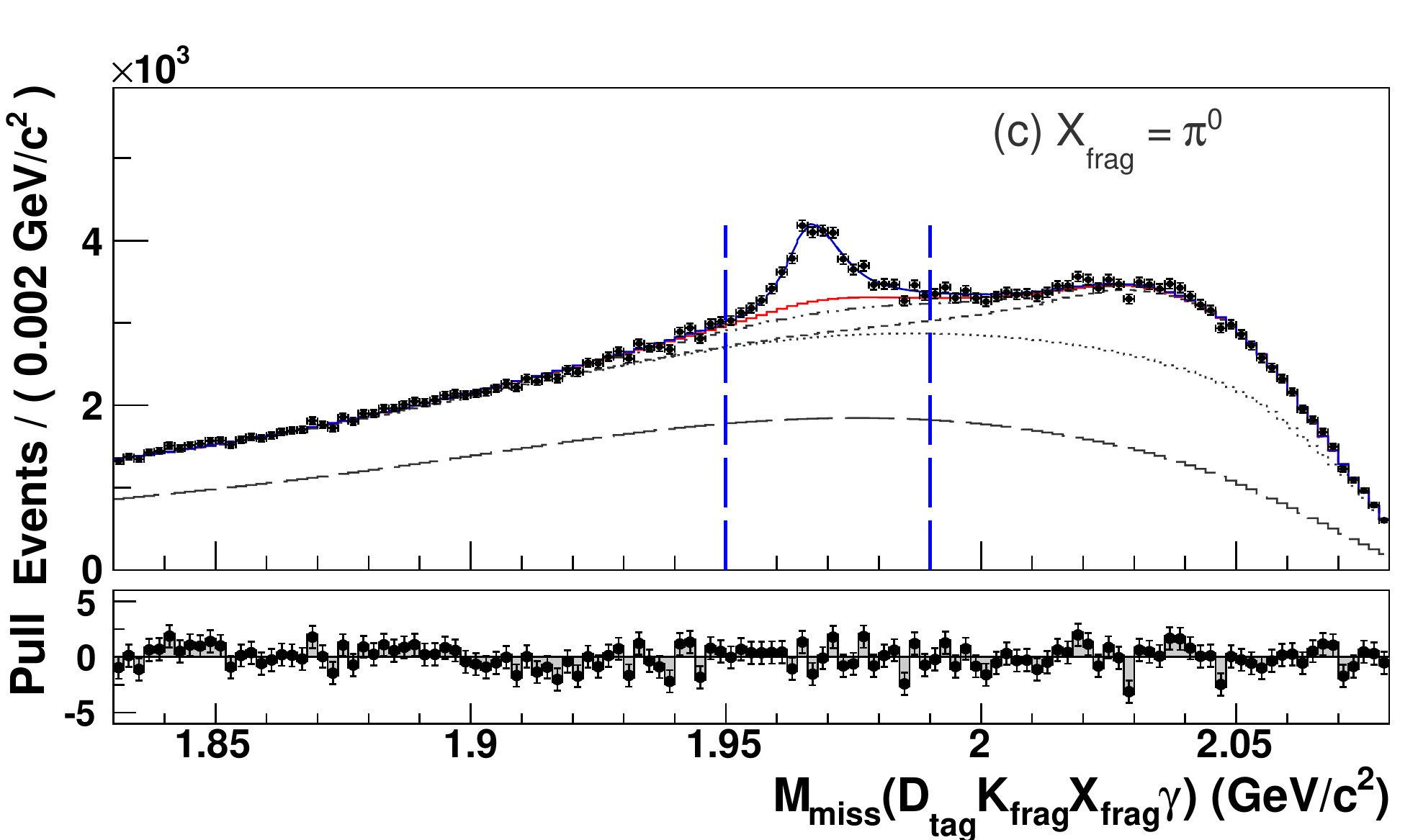}\\
 \includegraphics[width=0.475\textwidth]{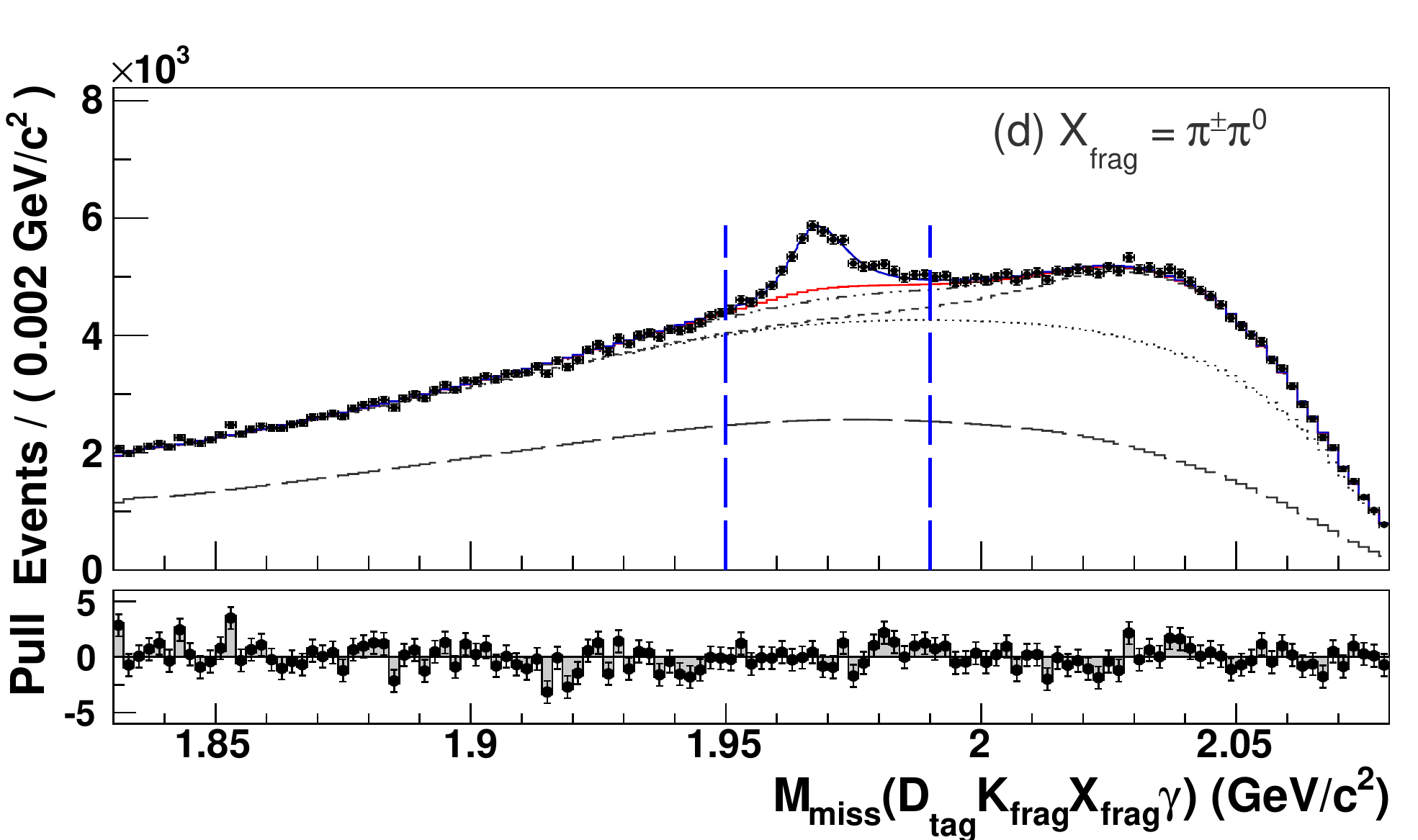}
 \includegraphics[width=0.475\textwidth]{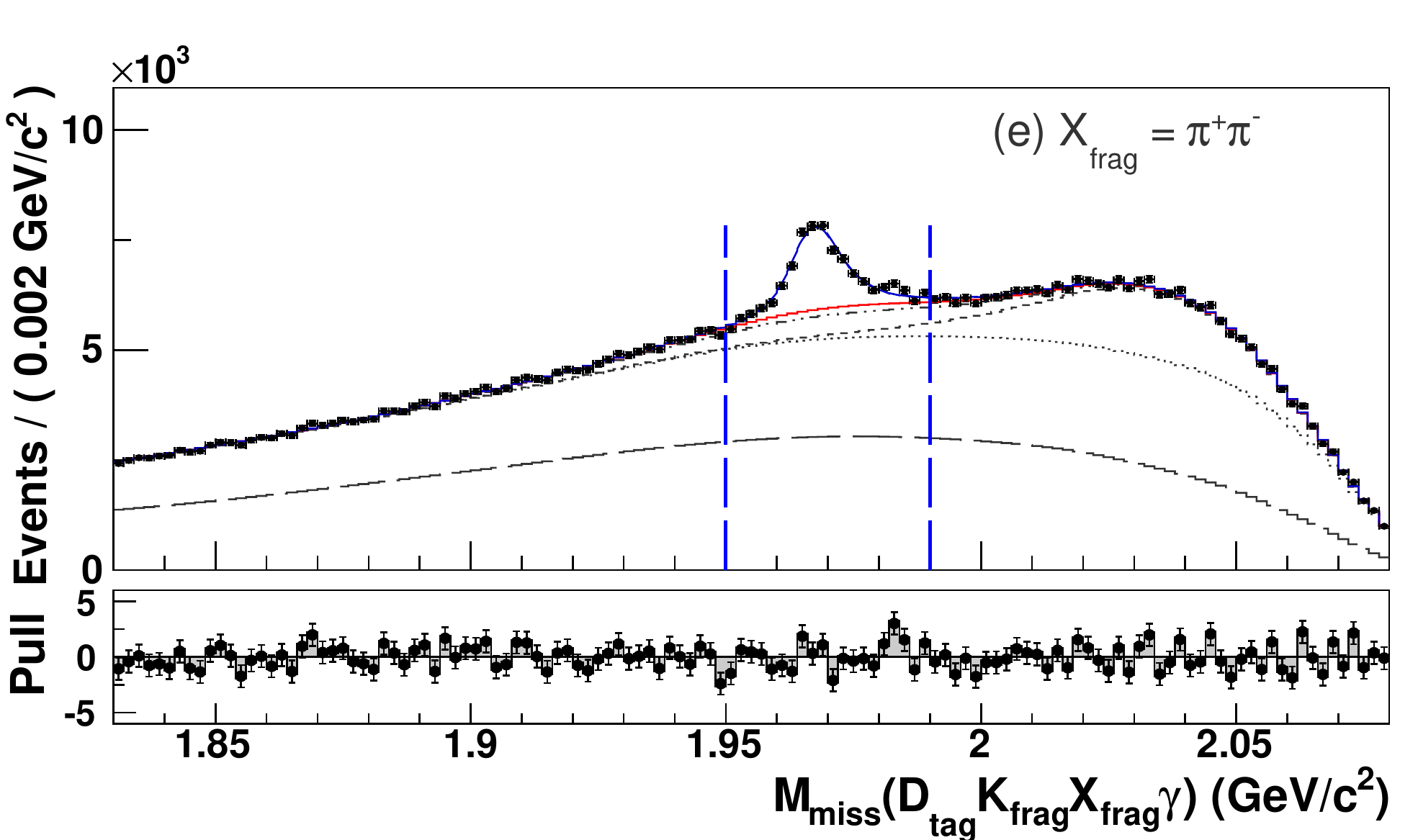}\\
 \includegraphics[width=0.475\textwidth]{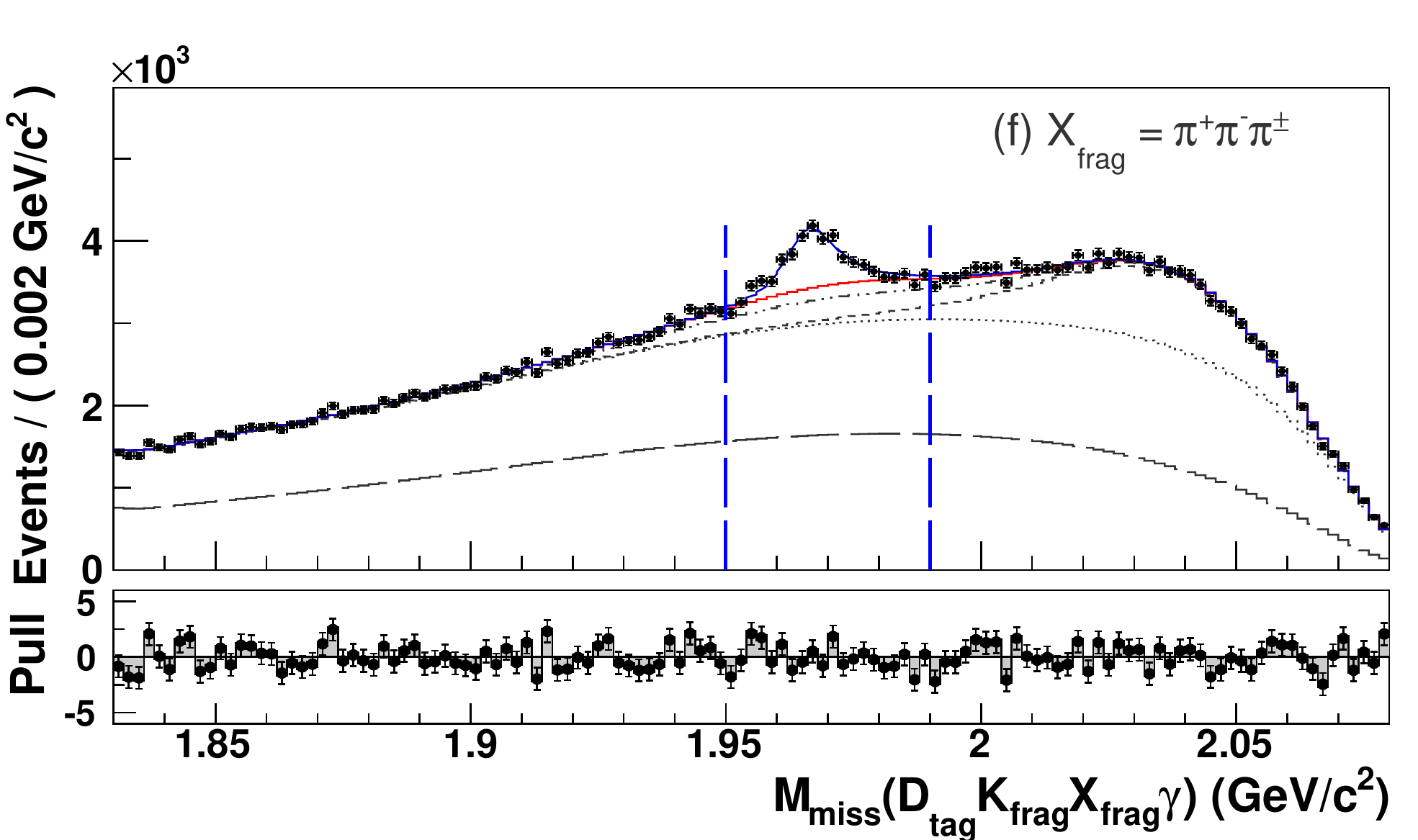}
 \includegraphics[width=0.475\textwidth]{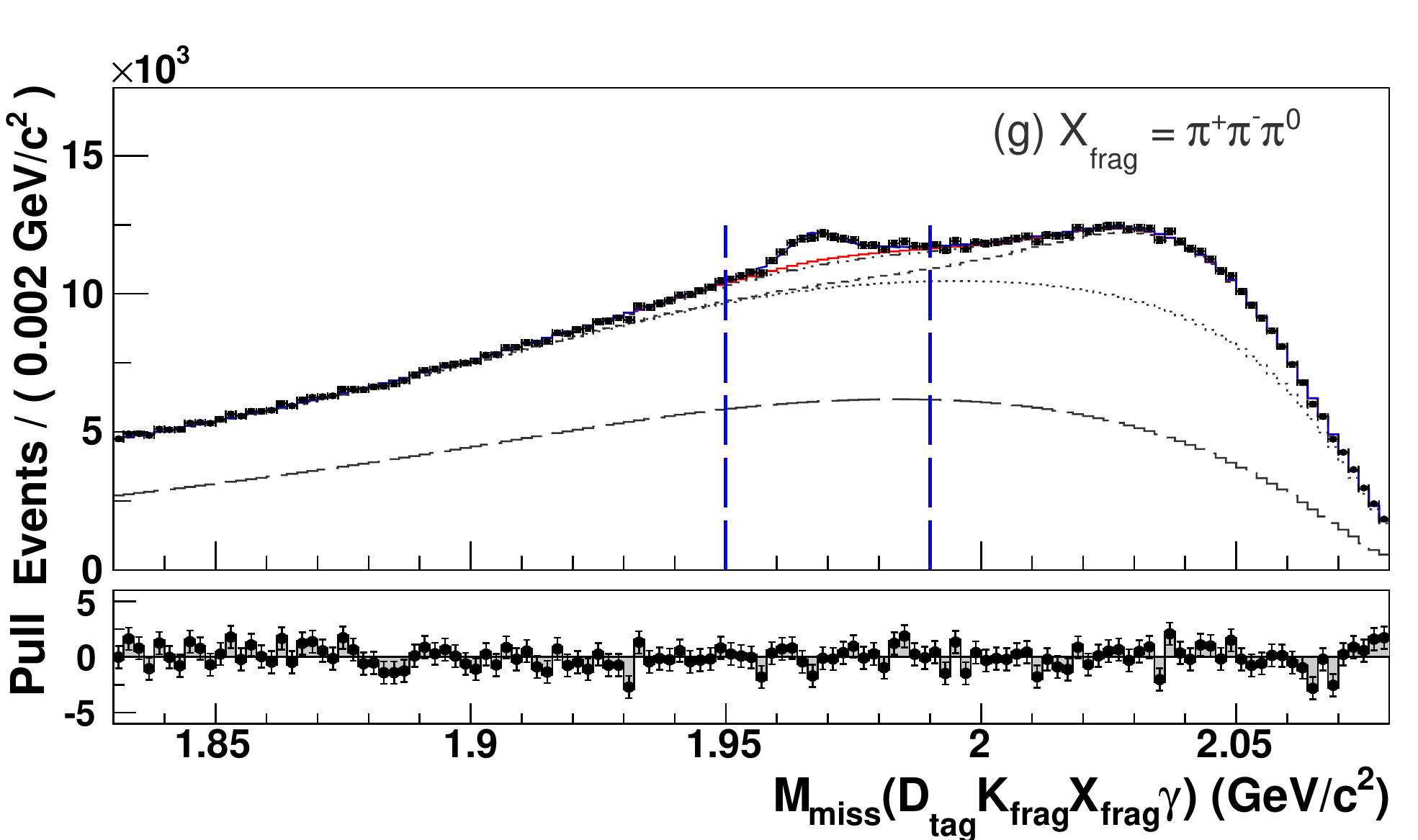}\\
 \caption{The $\mmiss(\dtagkx\gamma)$ distributions for all seven $\xfrag$ modes with superimposed fit results (solid blue line). 
 Within each panel, the curves show, from bottom to top, the cumulative contributions of background candidates where the signal 
 $\gamma$ originates from a $\pi^0$ that does not itself originate from a $D^{\ast}_{(s)}$ decay, background candidates with wrong $\gamma$, 
 background candidates where the signal $\gamma$ originates from a $\pi^0$ decay produced in $D_{(s)}^{\ast}\to D_{(s)}\pi^0$ decays,
 background candidates where the signal $\gamma$ candidate originates from a $D^{\ast0}\to D^0\gamma$ decays or
 mis-reconstructed signal candidates, and signal candidates.
 The two blue dashed vertical lines indicate the $\mmiss(\dtagkx\gamma)$ signal region. The normalized fit residual (referred in this
 and all other figures as ``Pull'') is defined as $(N_{\rm observed}-N_{\rm fit})/\sqrt{N_{\rm observed}}$.}
 \label{fig:dsincl:datafitres}
\end{figure}
The fitted inclusive $\dspm$ yields for each $\xfrag$ mode are given in table~\ref{tab:dsincl:yields}. The PDFs describe well the observed 
data distributions as can be seen from the normalized fit residuals in figure~\ref{fig:dsincl:datafitres}. The normalized $\chi^2$ values 
of the fits are between 1.06 and 1.32.
\begin{table}[t!]
 \centering
 \begin{tabular}{l|r@{\,$\pm$\,}l}
$\xfrag$ mode			& \multicolumn{2}{c}{$N_{D_s}^{\rm inc}$}  \\\hline\hline
nothing 			& $23460$&$280$ \\ 
$\pi^{\pm}$ 			& $23390$&$350$ \\ 
$\pi^0$ 			& $8030$&$480$  \\ 
$\pi^{\pm}\pi^0$ 		& $9290$&$550$  \\ 
$\pi^{\pm}\pi^{\mp}$ 		& $14930$&$450$ \\ 
$\pi^{\pm}\pi^{\mp}\pi^{\pm}$ 	& $5680$&$330$  \\ 
$\pi^{\pm}\pi^{\mp}\pi^0$ 	& $9580$&$820$  \\ 
\hline\hline
Sum 				& $94360$&$1310$ \\
 \end{tabular}
\caption{Yields of inclusively reconstructed $\dspm$ mesons per individual $\xfrag$ mode. The uncertainties are statistical only.}
\label{tab:dsincl:yields}
\end{table}

In the fit to the $\mmiss(\dtagkx\gamma)$ distributions, we set $\sigma_{\rm cal}=2.0$~\mevcc. To estimate the 
systematic uncertainty, we repeat the fits to the $\mmiss(\dtagkx\gamma)$ distributions by changing 
$\sigma_{\rm cal}$ by $\pm0.2$~\mevcc (one standard deviation) and assign the resulting difference of $\pm1.39\%$ in 
the inclusive $\dspm$ yield as the systematic error. In the nominal fit,
the fraction $f_{D^0\gamma/D_{(s)}\pi^0}$ is fixed to the MC-determined value. We vary this parameter by $\pm5\%$ to 
conservatively estimate the possible differences between data and MC in the relative production rates of $D^{\ast+}$ and 
$D^{\ast0}$ mesons. The impact on the inclusive $\dspm$ yields is found to be small ($\pm0.41\%$). To account for the limited statistics
of the MC sample used to determine the histogram PDFs, we repeatedly vary the contents of all bins of all histogram templates within 
their statistical uncertainties and refit. The systematic uncertainty is taken to be the root mean square (RMS) of
the obtained distribution of the inclusive signal yield ($\pm0.51\%$).

The total inclusive $\dspm$ yield, including systematic uncertainties, is 
\begin{equation}
 N^{\rm inc}_{\ds} = 94360\pm1310(\rm stat.)\pm1450(\rm syst.).
\label{eq:inclds:yield}
\end{equation}

\section{\boldmath Reconstruction of $\dsp$ decays within the inclusive $\dsp$ sample}
\label{sec:exclusiverec}
Using the inclusive sample of $\dsp$ mesons, we reconstruct specific $\dsp$ meson decays to the following final states: 
$K^-K^+\pi^+$, $\overline{K}{}^0K^+$, $\eta \pi^+$, $e^+\nu_{e}$, $\mu^+\nu_{\mu}$, and $\tau^+\nu_{\tau}$; for the last mode, the 
$\tau$ lepton is reconstructed via its decays to $e^+\nu_{e}\overline{\nu}{}_{\tau}$, $\mu^+\nu_{\mu}\overline{\nu}{}_{\tau}$, 
and $\pi^+\overline{\nu}{}_{\tau}$. We keep only inclusive $\dsp$ candidates within the signal region of $\mmiss(\dtagkx\gamma)$ defined as
$1.95~\mbox{\gevcc}<\mmiss(\dtagkx\gamma)<1.99$~\gevcc; the sideband regions are used to study background properties except in the case 
of the $\dskkpi$ decay, where all inclusive $\dsp$ candidates are kept. We find that 88.7\%  of correctly reconstructed inclusive 
$\dsp$ candidates populate the $\mmiss(\dtagkx\gamma)$ signal region, within which the purity of the inclusive $\dsp$ sample ranges 
between 3.5\% ($\xdmG$) and 41\% ($\xdmA$). 

In the following subsections, we describe the reconstruction procedure and signal yield extraction for the six decay modes.

\subsection{\boldmath $\ds^+\to K^-K^+\pi^+$}

The reconstruction of $\ds^+\to K^-K^+\pi^+$ decays requires exactly three charged tracks 
in the rest of the event with a net charge equal to the charge of the inclusively reconstructed $\dsp$ candidate. The track 
with charge opposite that of the inclusive $\ds^+$ candidate is selected to be the $K^-$ candidate while the two same-sign tracks 
are identified as $K^+$ or $\pi^+$ based on their likelihood ratios, ${\cal L}_{K,\pi}$.

The exclusively reconstructed  $\ds^+\to K^-K^+\pi^+$ candidates within the inclusive $\dsp$ sample are identified as a peak 
at the nominal mass of the $\ds^{\ast +}$ in the invariant mass distribution of the $K^-K^+\pi^+\gamma$ combination, 
$M(K^-K^+\pi^+\gamma)$. Here, $\gamma$ stands for the signal photon candidate used to reconstruct the inclusive $\dsp$ candidate in 
the recoil against the $\dtagkx\gamma$ system. The $M(K^-K^+\pi^+\gamma)$ is chosen over the $\dsp$ invariant mass, $M(K^-K^+\pi^+)$, because 
both sides --- inclusive and exclusive --- have to be correctly reconstructed to produce peaks in $M(K^-K^+\pi^+\gamma)$ and 
$\mmiss(\dtagkx\gamma)$. Correctly reconstructed $\ds^+\to K^-K^+\pi^+$ events will peak in $M(K^-K^+\pi^+)$ even if 
the inclusive reconstruction of $\dsp$ candidates fails, e.g., the photon candidate is incorrectly identified.
The $M(K^-K^+\pi^+\gamma)$ and $\mmiss(\dtagkx\gamma)$ are correlated due to their common input: the photon 
four-momentum. Imposing the requirement that inclusive $\dsp$ candidates populate the 
signal $\mmiss(\dtagkx\gamma)$ region would distort the background distribution to have a peaking structure
in $M(K^-K^+\pi^+\gamma)$. We avoid this by taking all inclusive $\dsp$ candidates into consideration rather than only
those populating the signal region in $\mmiss(\dtagkx\gamma)$. 

We parameterize the $M(K^-K^+\pi^+\gamma)$ distribution as 
\begin{eqnarray}
 {\cal F}(M(K^-K^+\pi^+\gamma)) & = & N_{\rm sig}\cdot {\cal H_{\rm sig}}(M(K^-K^+\pi^+\gamma)-\delta_M)\otimes {\cal G}(\sigma_{\rm cal}^{\rm excl})\nonumber\\
 & & {} + N_{\dsst\pi^0}\cdot {\cal H}_{\dsst\pi^0}(M(K^-K^+\pi^+\gamma))\nonumber\\
 & & {} + N_{\rm comb}\cdot \left[ 1+c_1\cdot M(K^-K^+\pi^+\gamma) \right. \nonumber\\&& \quad {} + \left. c_2\cdot M(K^-K^+\pi^+\gamma)^2+c_3\cdot M(K^-K^+\pi^+\gamma)^3 \right],
\end{eqnarray}
where the three terms describe correctly reconstructed $\ds^{\ast+}\to\dsp\gamma\to K^-K^+\pi^+\gamma$ decays (signal), 
mis-reconstructed $\ds^{\ast+}\to\dsp\pi^0\to K^-K^+\pi^+\gamma\gamma$ decays where one of the photons from the $\pi^0$ decay is lost,
and random combinations of charged tracks or photon (combinatorial background). The latter is parameterized as a third order polynomial while
the first two contributions are represented using the non-parametric histogram PDFs taken from MC. We convolve the signal histogram PDF
with a Gaussian function, ${\cal G}(\sigma_{\rm cal}^{\rm excl})$, to take into account the differences between the resolutions of 
$M(K^-K^+\pi^+\gamma)$ in real and MC samples. We estimate $\sigma_{\rm cal}^{\rm excl}$ by the procedure 
described in section \ref{sec:inclds:yield:extraction}, the only difference being that the resolution on $M(K^-K^+\pi^+\gamma)$ is calibrated 
instead of the $\ds^{\ast+}$ and $\dsp$ mass difference. We determine $\sigma_{\rm cal}^{\rm excl} = 3.2\pm0.2$~\mevcc. 
Free parameters of the fit are the normalization parameters, $N_i$, the position of the signal peak relative to the peak position
in the MC, $\delta_M$, and the combinatorial background shape parameters, $c_i$.

\begin{figure}[t!]
  \centering
  \includegraphics[width=1\textwidth]{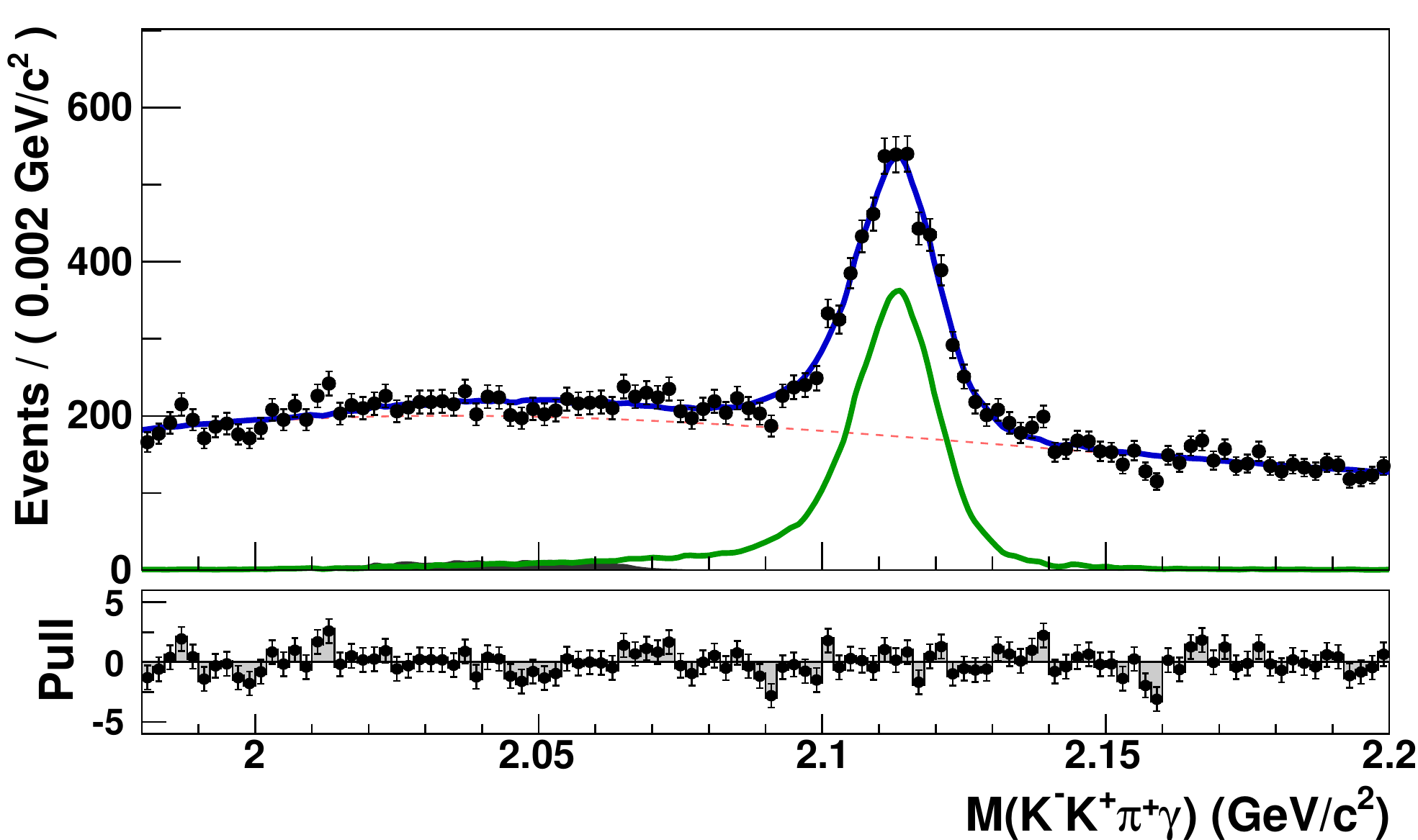} 
  \caption{The $M(K^-K^+\pi^+\gamma)$ distribution of exclusively reconstructed $\dsp\to K^-K^+\pi^+$ decays within the inclusive $\dsp$ sample 
  with superimposed fit results (solid blue line). The solid green line shows the signal contribution while the red dashed line shows the 
  contribution of combinatorial background; the contribution of $\dsp\to K^-K^+\pi^-$ candidates originating from 
  $\ds^{\ast+}\to\dsp\pi^0$ is indicated by the full dark gray histogram.}
  \label{figs:kkpi:daresults}
\end{figure}

The $M(K^-K^+\pi^+\gamma)$ distribution of exclusively reconstructed $\dsp\to K^-K^+\pi^+$ decays within the inclusive $\dsp$ sample 
is shown in figure \ref{figs:kkpi:daresults} with the superimposed fit. The number of correctly reconstructed 
$\dsp\to K^-K^+\pi^+$ decays is
\begin{equation}
 N(\dsp\to K^-K^+\pi^+)=4094\pm123,
 \label{eq:dskkpi:yieldDA}
\end{equation}
where the error is statistical only.

\subsection{\boldmath $\ds^+\to \overline{K}{}^0K^+$}

We reconstruct $\ds^+\to \overline{K}{}^0K^+$ decays by requiring only one additional charged kaon in the rest of the event
whose charge equals that of the inclusively reconstructed $\dsp$ candidate.
The neutral kaon is not reconstructed;
rather, it is identified as a peak at the nominal mass-squared of the neutral kaon in the missing-mass-squared distribution
\[
 \mmiss^2(\dtagkx\gamma K) = p_{\rm miss}^2(\dtagkx\gamma K),
\]
where the missing four-momentum is given by
\[
 p_{\rm miss}(\dtagkx\gamma K)= p_{e^+} + p_{e^-} - p_{\dtag} - p_{K_{\rm frag}} - p_{\xfrag} - p_{\gamma} - p_{K}.
\]
An explicit reconstruction of the $\overline{K}{}^0$ meson to two oppositely charged pions would lead to a significant signal loss.
The signal peak of $\ds^+\to \overline{K}{}^0K^+$ in the $\mmiss^2(\dtagkx\gamma K)$ distribution is 
used to calibrate the $\mmiss^2$ resolution, which is important in the extraction of the signal yield of $\dsmunu$ decays.
Since the flavor of the neutral kaon is not determined, the doubly Cabibbo suppressed decays, $\ds^+\to K^0K^+$, 
also contribute to the peak in $\mmiss^2(\dtagkx\gamma K)$. Their relative contribution can be estimated naively
to be equal to $\tan^4\theta_C\approx 0.29\%$ ($\theta_C$ being the Cabibbo mixing angle), which is an order of 
magnitude below the expected statistical uncertainty and thus safely neglected.

The signal yield of partially reconstructed $\ds^+\to \overline{K}{}^0K^+$ decays is extracted by performing a binned extended 
maximum likelihood fit to the $\mmiss^2(\dtagkx\gamma K)$ distribution. The signal component is parameterized as a sum of three 
Gaussian functions with a common mean. In the fit, we fix the signal shape parameters to the values determined from the MC sample except 
for the mean and the resolution scaling factor, $s$, of the core and the second Gaussian function. In addition to the signal contribution,
two-body $\dsp\to \pi^0K^+$ and $\dsp\to \eta K^+$ decays peak in the $\mmiss^2(\dtagkx\gamma K)$ distribution. We use the same 
parameterization for these peaking backgrounds as for signal except that we fix their means to the nominal masses squared of 
$\pi^0$ and $\eta$ mesons. Other background sources that produce distinct structures in $\mmiss^2(\dtagkx\gamma K)$ are: 
$\ds\to K^{\ast +} \overline{K}{}^0\to K^+\pi^0 \overline{K}{}^0$ decays and $\ds^+\to \eta \pi^+$ decays where the $\pi^+$ is 
misidentified as signal $K^+$ candidate. Both contributions are parameterized as the sum of a Gaussian and a bifurcated Gaussian 
function, where all parameters are fixed to the values determined from the MC sample. The combinatorial background is parameterized 
with a fourth-order polynomial whose coefficients are determined with the fit to the $\mmiss^2(\dtagkx\gamma K)$ distribution 
for inclusive $\dsp$ candidates in the $\mmiss(\dtagkx\gamma)<1.95$~\gevcc sideband region. Yields of all but two event categories 
are free parameters of the fit; the $\dsp\to \eta K^+$ and $\dsp\to \eta \pi^+$ yields are constrained to the expected values based on 
their known branching fractions and MC determined efficiencies.

\begin{figure}[t]
 \centering
 \includegraphics[width=1\textwidth]{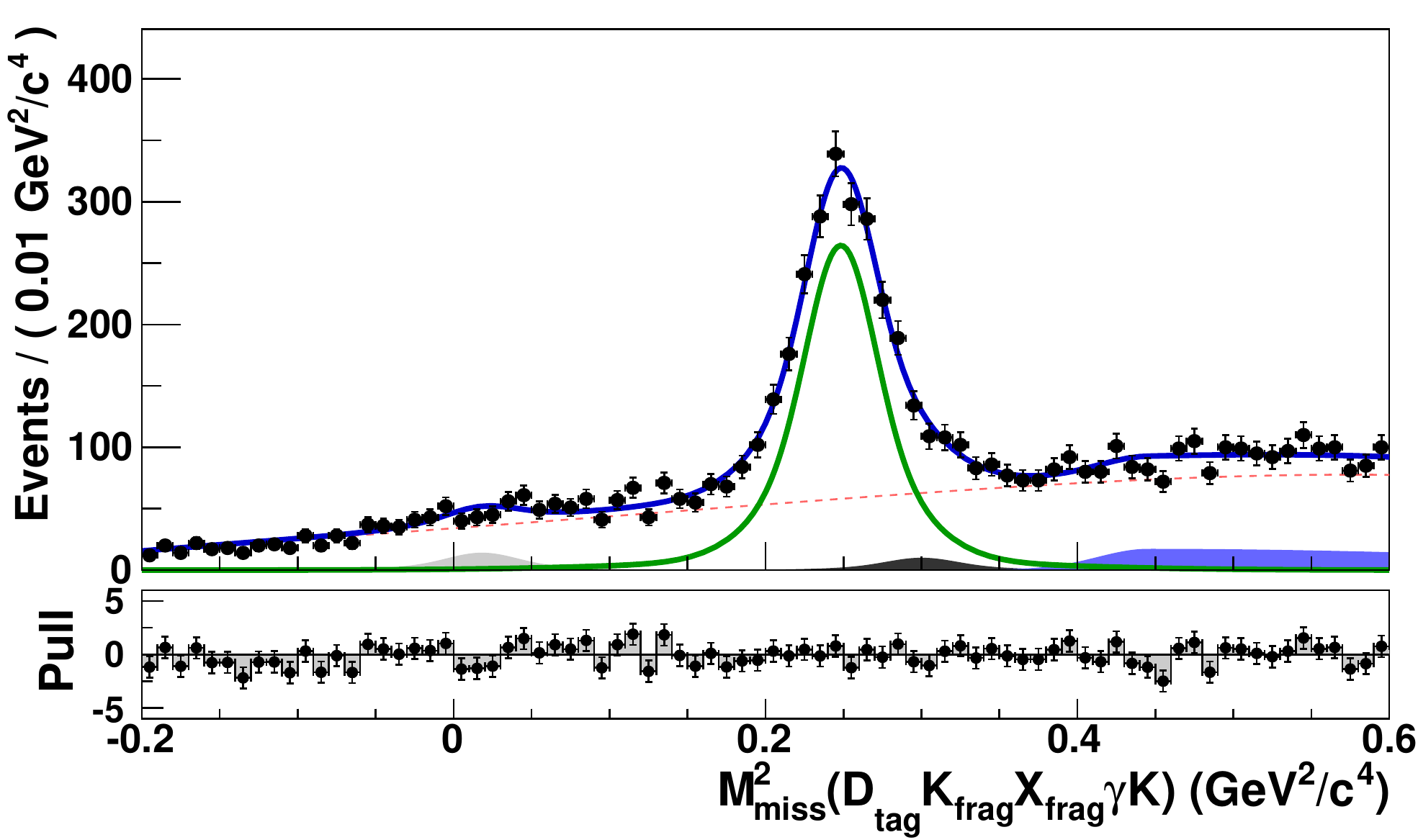} 
 \caption{The $\mmiss^2(\dtagkx\gamma K)$  distribution of partially reconstructed $\ds^+\to \overline{K}{}^0K^+$ decays within 
the inclusive $\dsp$ sample with superimposed fit results (solid blue line). The solid green line shows the signal contribution, 
the dashed red line the contribution of combinatorial background, while the full histograms show the contributions of
$\ds^+\to \pi^0 K^+$ (light gray), $\ds^+\to \eta K^+$ (dark gray) or $\ds^+\to K^{\ast+}\overline{K}{}^0$ decays (blue).}
\label{figs:kk0:daresults}
\end{figure}

The  $\mmiss^2(\dtagkx\gamma K)$ distribution of partially reconstructed  $\ds^+\to \overline{K}{}^0K^+$ decays 
within the inclusive $\dsp$ sample is shown in figure \ref{figs:kk0:daresults} with the superimposed fit. 
The number of correctly reconstructed $\ds^+\to \overline{K}{}^0K^+$ decays is
\begin{equation}
 N(\dsp\to \overline{K}{}^0 K^+)=2018\pm75,
 \label{eq:dskk0:yieldDA}
\end{equation}
where the error is statistical only. The yield of Cabibbo suppressed $\dsp\to \pi^0K^+$ decays, $108\pm 31$ (statistical error only), 
is found to be consistent within uncertainties with the expectation, $52\pm 18$, based on a measurement performed by CLEO \cite{Mendez:2009aa}.

\subsection{\boldmath $\ds^+\to \eta \pi^+$}

As in the case of $\ds^+\to \overline{K}{}^0K^+$ decays, we perform a partial reconstruction of $\ds^+\to \eta \pi^+$ decays. We require 
only one charged track consistent with the pion hypothesis in the rest of the event. To avoid a significant signal loss, we do not perform an 
explicit reconstruction of the $\eta$ meson but rather identify it as a peak at the nominal mass-squared of the $\eta$ in the 
missing-mass-squared distribution
\[
 \mmiss^2(\dtagkx\gamma \pi) = p_{\rm miss}^2(\dtagkx\gamma \pi),
\]
where the missing four-momentum is given by
\[
 p_{\rm miss}(\dtagkx\gamma \pi)= p_{e^+} + p_{e^-} - p_{\dtag} - p_{K_{\rm frag}} - p_{\xfrag} - p_{\gamma} - p_{\pi}.
\]

The sample of events with inclusive $\dsp$ candidate plus one additional positively charged pion contains a significant contribution of 
$\dsp\to\tau^+\nu_{\tau}$ decays with the tau lepton decaying hadronically to a charged pion and a neutrino. The contribution of these 
events is suppressed by requiring that the extra neutral energy in the ECL ($E_{\rm ECL}$) be larger than 1.0~\gev, 
where $E_{\rm ECL}$ represents the sum over all energy deposits in the ECL that are not associated with the charged pion candidate 
and the tracks and neutrals used in the inclusive reconstruction of the $\dsp$ candidate~\cite{Adachi:2012mm}. 
The $\dsp\to\taunu\to\pi^+\overline{\nu}{}_{\tau}\nu_{\tau}$ decays peak at zero in $E_{\rm ECL}$ while $\dsetapi$ 
decays deposit a significant amount of energy in the ECL via the $\eta$ decay products. (See section \ref{sec:dstaunu} for more details.)

The signal yield of partially reconstructed $\ds^+\to \eta \pi^+$ decays is extracted by performing a binned maximum likelihood fit to the 
$\mmiss^2(\dtagkx\gamma \pi)$ distribution. We use the same parameterization for the signal component as in the case of 
$\dskzk$ decays. In addition, the resolution scaling factor of the core and the second Gaussian, $s$, is allowed to float within 
a Gaussian constraint in the fit, where the mean and width of the constraint are set to the value and uncertainty determined in the 
fit of the $\dskzk$ candidates ($s=1.177\pm0.052$), respectively. We identify a single source of peaking background to be the two-body 
$\ds^+\to K^0\pi^+$ decay that peaks at the neutral kaon mass-squared and is parameterized in the same way as the signal. Other 
background sources that produce distinct structures in $\mmiss^2(\dtagkx\gamma \pi)$ are $\ds^+\to \overline{K}{}^0 K^+$
decays with a misidentified kaon and $\dsp\to \rho^0 K^+\to \pi^+\pi^-K^+$ decays; both are parameterized
as bifurcated Gaussian functions. The combinatorial background is parameterized with a fourth order polynomial whose
coefficients are fixed to the values determined with the fit to the $\mmiss^2(\dtagkx\gamma \pi)$ distribution 
of inclusive $\dsp$ candidates from the $\mmiss(\dtagkx\gamma)<1.95$~\gevcc sideband region. Yields of all but two 
event categories are free parameters of the fit; the $\dsp\to K^0\pi^+$ and $\ds^+\to \overline{K}{}^0 K^+$ yields 
are constrained to expected values based on their known branching fractions and MC-determined efficiencies.
\begin{figure}[t!]
 \centering
 \includegraphics[width=1\textwidth]{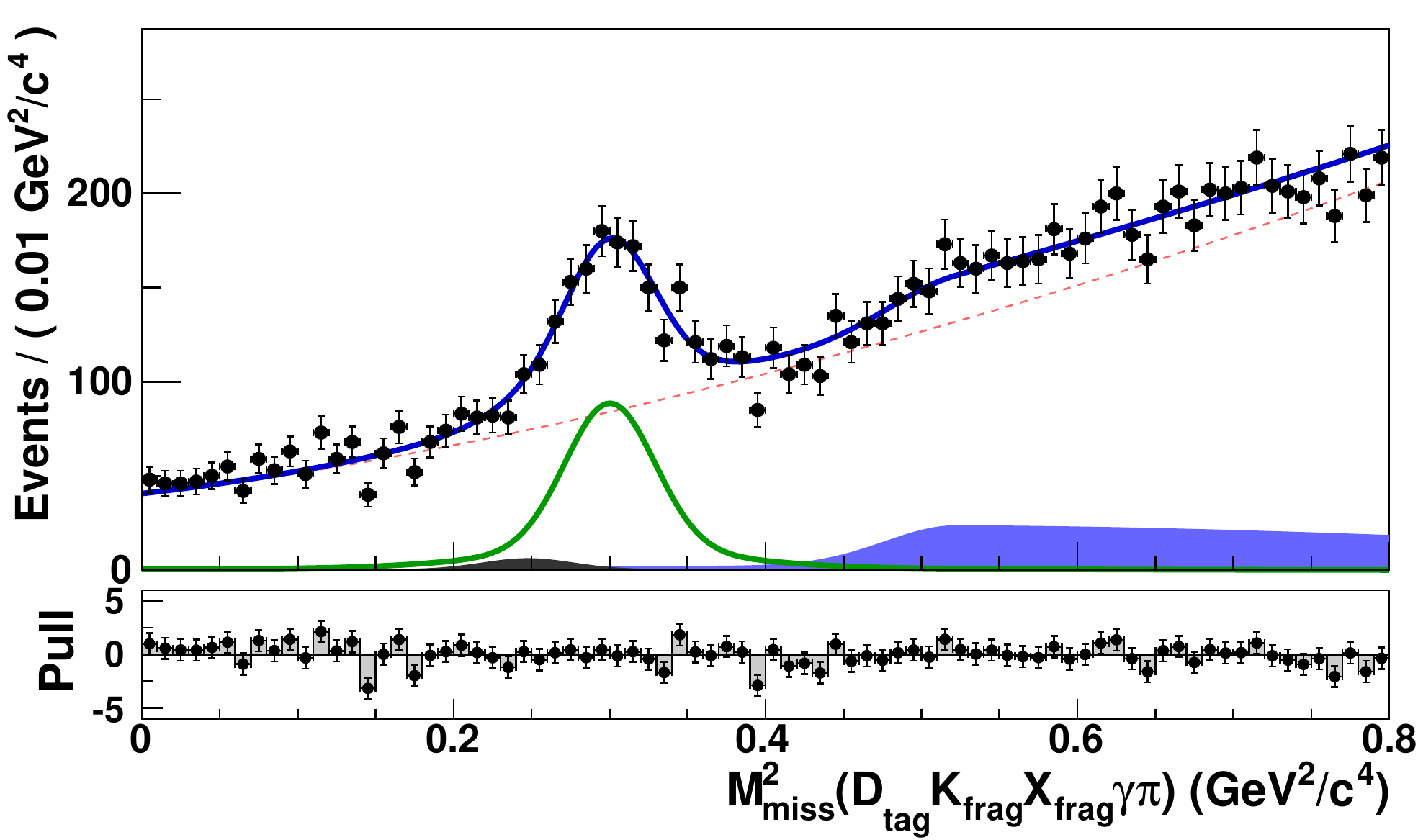} 
 \caption{The $\mmiss^2(\dtagkx\gamma \pi)$  distribution of partially reconstructed $\ds^+\to \eta \pi^+$ decays within 
the inclusive $\dsp$ sample with superimposed fit results (solid blue line). Solid green line shows the signal contribution,
dashed red line contribution of combinatorial background, while the full histograms 
show the contributions of $\ds^+\to K^0 \pi^+$ (dark gray) or $\ds^+\to\overline{K}{}^0 K^+$ and $\dsp\to \rho^0 K^+$ decays (blue).}
 \label{figs:etapi:daresults}
\end{figure}

The $\mmiss^2(\dtagkx\gamma \pi)$ distribution is shown in figure \ref{figs:etapi:daresults} with the superimposed fit. 
The number of correctly reconstructed $\ds^+\to \eta\pi^+$decays is 
\begin{equation}
 N(\dsetapi)=788\pm59,
 \label{eq:dsetapi:yieldDA}
\end{equation}
where the error is statistical only.

\subsection{\boldmath $\dsmunu$}
The $\dsmunu$ decays are reconstructed by requiring one additional charged track consistent with the muon hypothesis
in the rest of the event. The single missing neutrino is then identified as a peak at zero in the missing-mass-squared distribution
\[
 \mmiss^2(\dtagkx\gamma \mu) = p_{\rm miss}^2(\dtagkx\gamma \mu),
\]
where the missing four-momentum is given by
\[
 p_{\rm miss}(\dtagkx\gamma \mu)= p_{e^+} + p_{e^-} - p_{\dtag} - p_{K_{\rm frag}} - p_{\xfrag} - p_{\gamma} - p_{\mu}.
\]

The signal yield is extracted by performing a binned maximum likelihood fit to the $\mmiss^2(\dtagkx\gamma \mu)$ distribution. The 
signal component is parameterized as the sum of three Gaussian functions with a common mean, where all parameters except the mean 
are fixed to their MC-determined values. As in the case of $\dsetapi$ decays, the resolution scaling factor of the first and 
second Gaussians, $s$, is allowed to float within a Gaussian constraint in the fit, except that the mean and width of the 
constraint here are set to the value and uncertainty determined in the fits of the $\dskzk$ and $\dsetapi$ candidates ($s=1.177\pm0.049$). 
The leptonic $\ds^+\to\taunu\to \mu^+\nu_{\mu}\nu_{\tau}\overline{\nu}{}_{\tau}$ decays produce three neutrinos in the final 
state and therefore do not peak at $\mmiss^2(\dtagkx\gamma \mu)$. Their contribution is found to be very well described by an exponential
function. In the fit, we include also contributions of hadronic $\dsetapi$ and $\dskzk$ decays where the muon candidate is
a misidentified pion or kaon, respectively. The former is parameterized as the sum of two Gaussian functions and the latter as a bifurcated
Gaussian function. In both cases, we fix the shape parameters to the MC-determined values. The combinatorial background is 
parameterized with an exponential function whose shape parameter is fixed to the value determined from the fit to 
the $\mmiss^2(\dtagkx\gamma \mu)$ distribution for candidates in the $\mmiss(\dtagkx\gamma)<1.95$~\gevcc sideband region. Free parameters
of the fits are the yield parameters of all but two spectral components; the $\dsetapi$ and $\dskzk$ yields are constrained 
to the expected values based on their measured branching fractions and MC-determined efficiencies. 

\begin{figure}[t]
 \centering
 \includegraphics[width=1\textwidth]{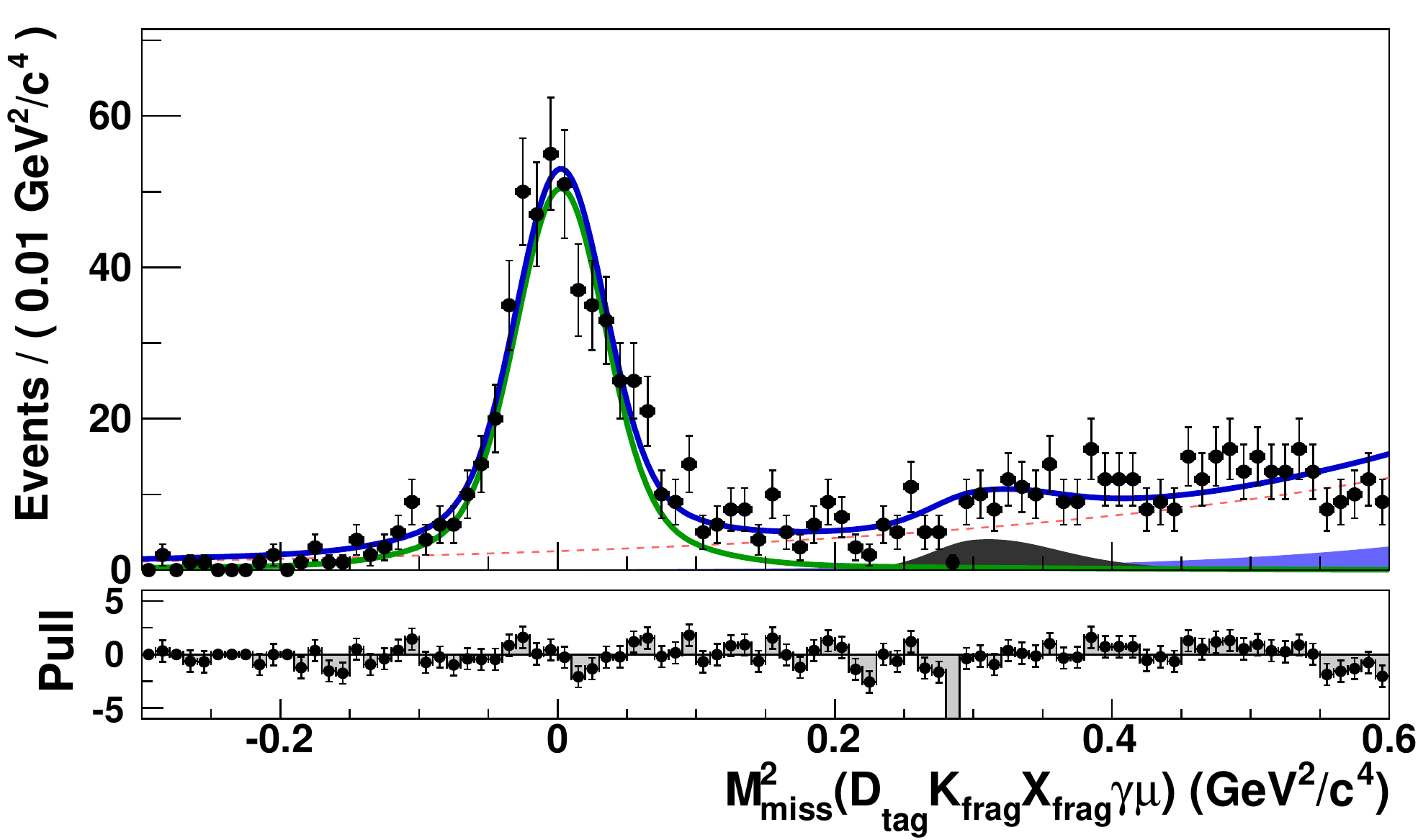} 
 \caption{The $\mmiss^2(\dtagkx\gamma \mu)$ distribution of exclusively reconstructed $\dsmunu$
 decays within the inclusive $\dsp$ sample superimposed fit results (solid blue line). The solid green line shows the contribution
 of signal, the red dashed line the contribution of combinatorial background, while the contributions of $\dsp\to\taunu$ and $\dsp\to \kzk$ or 
 $\eta\pi^+$ decays are indicated by the full blue and dark gray histograms, respectively.}
 \label{figs:munu:dataresults}
\end{figure}
The distribution of $\mmiss(\dtagkx\gamma\mu)$ with superimposed fit is shown in figure \ref{figs:munu:dataresults}. 
The number of reconstructed $\dsmunu$ decays is 
\begin{equation}
   N(\dsmunu)=492\pm26,
  \label{eq:dsmunu:yieldDA}
\end{equation}
where the error is statistical only. 

\subsection{\boldmath $\ds^+\to \tau^+\nu_{\tau}$}  
\label{sec:dstaunu}
The reconstruction of $\ds^+\to \tau^+\nu_{\tau}$ requires one charged track in the rest of the event that is identified as an electron, 
muon or a pion (denoted as $\dsp\to\tau^+(X^+)\nu_{\tau}$ where $X^+=e^+,~\mu^+$ or $~\pi^+$) indicating the subsequent decay of the 
$\tau^+$ lepton to $e^+\nu_e\overline{\nu}{}_{\tau}$, $\mu^+\nu_{\mu}\overline{\nu}{}_{\tau}$ or $\pi^+\overline{\nu}{}_{\tau}$.\footnote{The 
three decay modes cover almost half of all possible tau decays.}
Due to the multiple neutrinos in the final state, these decays do not peak in the missing-mass-squared distribution:
\[
 \mmiss^2(\dtagkx\gamma X) = p_{\rm miss}^2(\dtagkx\gamma X),
\]
where the missing four-momentum is given by
\[
 p_{\rm miss}(\dtagkx\gamma X)= p_{e^+} + p_{e^-} - p_{\dtag} - p_{K_{\rm frag}} - p_{\xfrag} - p_{\gamma} - p_{X}.
\]

The background in the $\dsp\to\tau^+(\pi^+)\nu_{\tau}$ sample is much larger than in the leptonic modes, but is reduced significantly 
by requiring the magnitude of the missing momentum of the event, $|\vec{p}_{\rm miss}(\dtagkx\gamma \pi)|$, to 
be larger than 1.2~\gevc in the laboratory frame. The background in this sample is further reduced by requiring 
$0.0<\mmiss^2(\dtagkx\gamma \pi)<0.6$~\gevsqcq.\footnote{Due to 
the lack of phase space and the fact that $\dsp\to\taunu\to\pi^+\overline{\nu}{}_{\tau}\nu_{\tau}$ decays have only two neutrinos in the final 
state, these decays populate a relatively narrow region in $\mmiss^2(\dtagkx\gamma \pi)$.} In the case of 
$\dsp\to\taunu\to\ell^+\nu_{\ell}\overline{\nu}{}_{\tau}\nu_{\tau}$ (where $\ell^+ = e^+$ or $\mu^+$), we require 
$\mmiss^2(\dtagkx\gamma \ell)>0.3$~\gevsqcq to veto $\ds^+\to\ell^+\nu_{\ell}$ decays. 

The signal yield of $\dsp\to \taunu$ decays is extracted from the simultaneous binned maximum likelihood fit to the $\eecl$ distributions of the 
three tau submodes. The signal decay has either zero or a small value of $\eecl$, while background events tend to have larger 
values due to the contributions from additional neutral clusters. Signal components include the cross-feed contribution from 
other $\tau$ decays: in the case of leptonic $\tau$ decays, the cross-feed contribution is found to be small (around 3\% of the
signal contribution from leptonic modes) while, in the case of hadronic $\tau^+\to\pi^+\overline{\nu}{}_{\tau}$ decays, a large 
cross-feed contribution originates from hadronic $\tau^+\to\rho^+\overline{\nu}{}_{\tau}$ decays by missing the neutral pion 
from the $\rho^+$ decay (20\% of the signal contribution from the pion mode). 
Backgrounds from several different $\dsp$ decays are found to contribute to $\dsp\to \tau^+(X^+)\nu_{\tau}$ samples 
and are listed in table~\ref{tab:taunu:bkgest}. The peaking background is dominated by $\dsp$ decays with $K^0_L$ in the final state, e.g., $\dsp\to \overline{K}{}^0\ell^+\nu_{\ell}$ 
in the case of leptonic tau decays or $\dskzk$ in the case of hadronic tau decays. If the $K^0_L$ deposits 
little or no energy in the ECL then these decay modes produce an $\eecl$ distribution very similar to the signal. The non-peaking background,
dominated by inclusive $\eta$ decays, is much less problematic since it rises smoothly with increasing $\eecl$. The $\eecl$ distributions of the above categories as well as of the combinatorial background are described 
with non-parametric histogram PDFs taken from MC samples. In the final fit, four parameters are allowed to vary: the total signal yield
summed over three tau decay modes (we fix the relative contribution of the signal from each tau decay mode $i$ to the total signal yield 
with the ratio of $(\br(\tau\to i)\cdot\varepsilon_i)/(\sum_{i=1}^3\br(\tau\to i)\cdot\varepsilon_i)$) and the yields of 
combinatorial background in each tau decay mode. The background contributions from $\dsp$ decays are fixed to the values given in table~\ref{tab:taunu:bkgest}. 
\begin{table}[t!]
\centering
\begin{tabular}{lr@{\,$\pm$\,}lr@{\,$\pm$\,}lr@{\,$\pm$\,}l}\hline\hline
					& \multicolumn{6}{c}{Estimated background yields}\\
Background Source 			& \multicolumn{2}{c}{$\tau^+(e^+)\nu_{\tau}$}	& \multicolumn{2}{c}{$\tau^+(\mu^+)\nu_{\tau}$}	& \multicolumn{2}{c}{$\tau^+(\pi^+)\nu_{\tau}$}\\ \hline\hline
$\dsp\to\eta\ell^+\nu_{\ell} $		& $911.0$&$102.3$		& $768.7$&$86.4$			& \multicolumn{2}{c}{--}\\
$\dsp\to\eta'\ell^+\nu_{\ell}$		& $49.5$&$12.0$			& $35.1$&$8.6$				& \multicolumn{2}{c}{--}\\
$\dsp\to\phi\ell^+\nu_{\ell} $		& $307.8$&$20.7$		& $188.0$&$13.3$			& \multicolumn{2}{c}{--}\\
$\dsp\to \overline{K}{}^0\ell^+\nu_{\ell}$		& $242.6$&$66.3$		& $175.7$&$48.1$			& \multicolumn{2}{c}{--}\\
$\dsp\to \overline{K}{}^{\ast0}\ell^+\nu_{\ell}$	& $26.0$&$10.5$			& $13.9$&$5.8$				& \multicolumn{2}{c}{--}\\
$\dsp\to K\overline{K}\ell^+\nu_{\ell}$	& $59.2$&$14.5$			& $33.1$&$8.0$				& \multicolumn{2}{c}{--}\\
$\dsp\to\mu^+\nu_{\mu}$			& \multicolumn{2}{c}{--}	& $10.0$&$1.4$				& $26.2$&$3.7$\\
$\dskzk$				& $18.5$&$2.5$			& $40.5$&$4.9$				& $132.3$&$9.2$\\
$\dsp\to \phi\pi^+$			& $11.2$&$2.1$			& $14.8$&$2.5$				& \multicolumn{2}{c}{--}\\
$\dsp\to K^{\ast+}K^0$   		& $32.4$&$8.3$			& $41.7$&$10.6$				& \multicolumn{2}{c}{--}\\
$\dsp\to \eta\pi^+$			& \multicolumn{2}{c}{--}	& \multicolumn{2}{c}{--}		& $398.2$&$24.2$\\
$\dsp\to \rho^0 K^+$	   		& \multicolumn{2}{c}{--}	& \multicolumn{2}{c}{--}		& $185.1$&$34.9$\\
\hline\hline
\end{tabular}
\caption{Estimated background yields of various $\dsp$ decays contributing to the three $\ds\to\taunu$ samples. The uncertainties include
the uncertainty of their branching fractions (taken from \cite{Beringer:1900zz}), the error on the inclusive $\dsp$ yield, as well as 
the uncertainty due to limited MC sample size used to determine the efficiencies and systematic uncertainty related to 
particle identification.}
\label{tab:taunu:bkgest}
\end{table}

\begin{figure}[t!]
 \centering
 \includegraphics[width=0.68\textwidth]{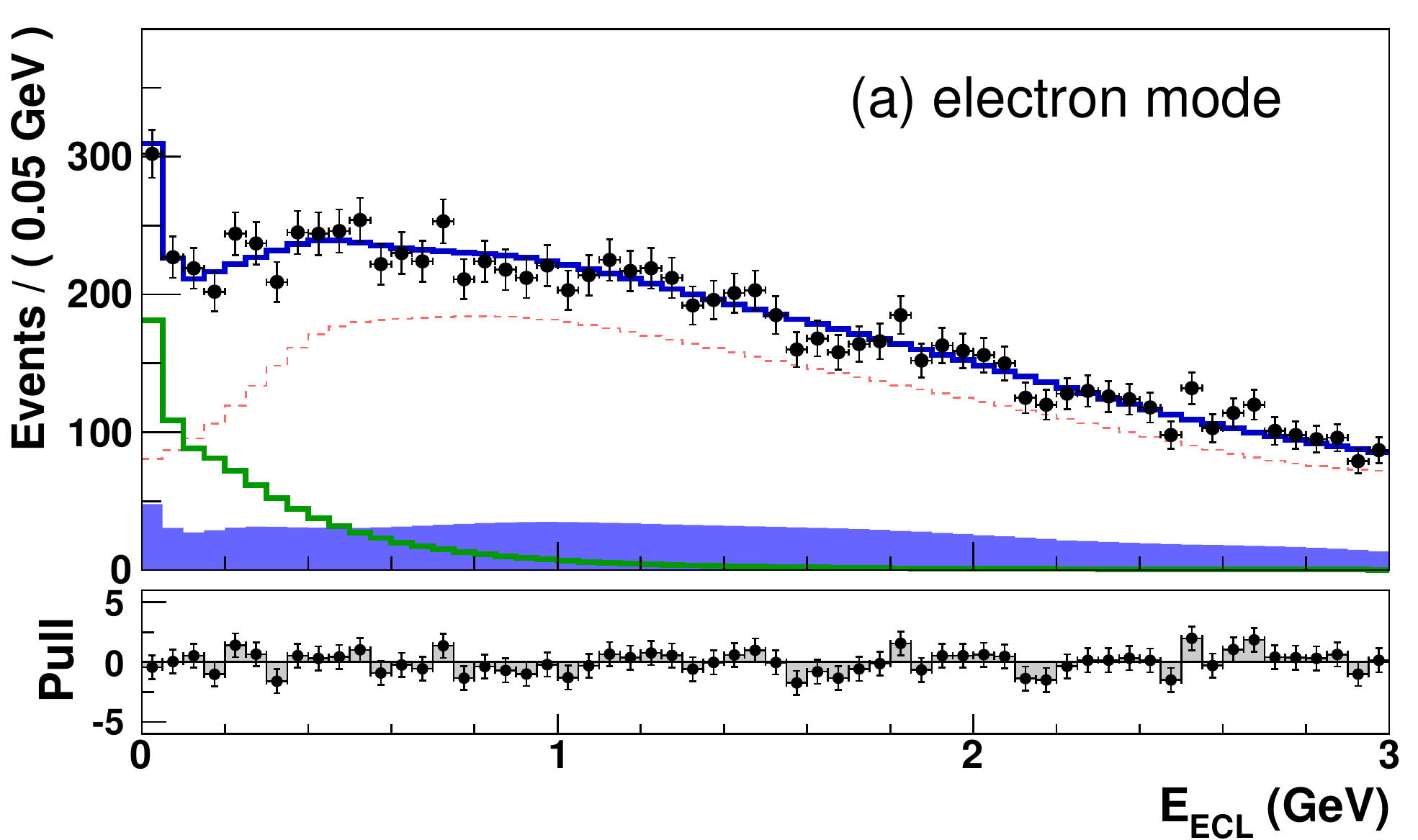}
 \includegraphics[width=0.68\textwidth]{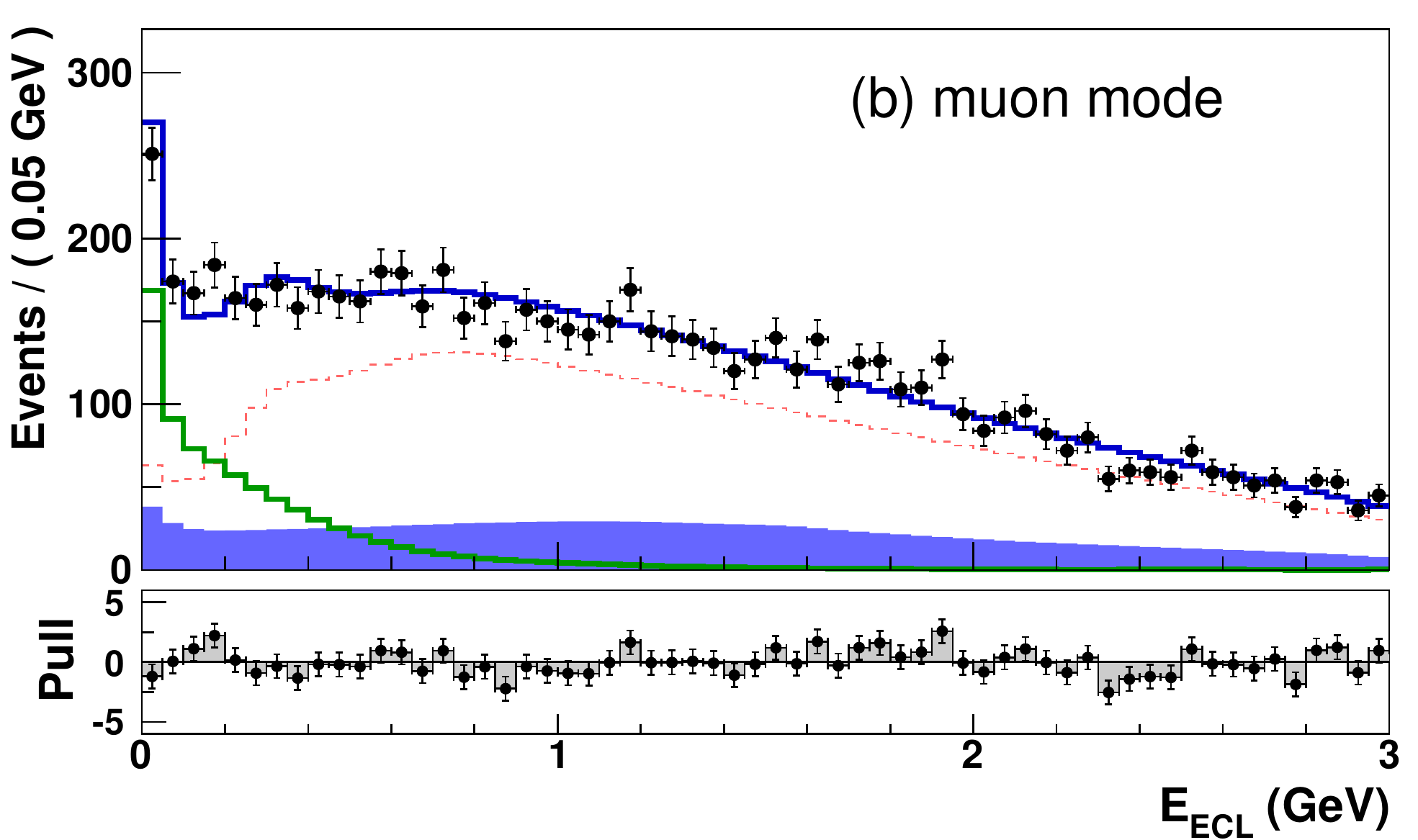}\\  
 \includegraphics[width=0.68\textwidth]{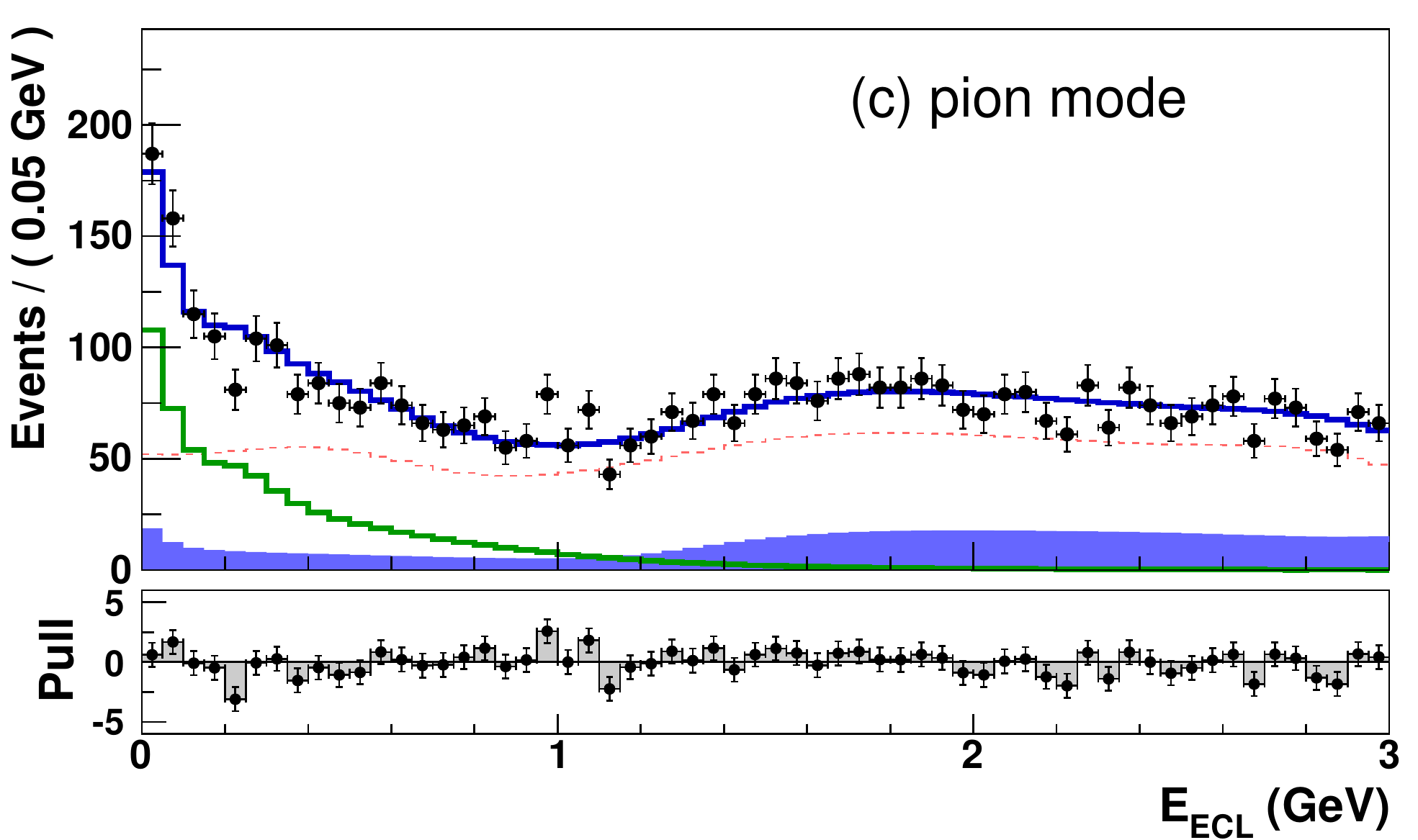}\\ 
 \caption{The $\eecl$ distribution of exclusively reconstructed $\dsp\to\tauenu$ (a), $\dsp\to\taumunu$ (b), and $\dsp\to\taupinu$ (c) 
 decays within the inclusive $\dsp$ sample with superimposed fit results. The solid green lines show the contributions of signal and 
 $\tau$ cross-feed, the red dashed line the contributions of combinatorial background, while the contributions of
 background from other $\dsp\to f$ decays are shown by the full blue histogram.}
 \label{figs:taunu:dataresults}
\end{figure}
Figure~\ref{figs:taunu:dataresults} shows the result of the simultaneous fit to the three $E_{\rm ECL}$ distributions for 
$\dsp\to\tauenu$, $\dsp\to\taumunu$, and $\dsp\to\taupinu$ decays. The signal yield is 
\begin{equation}
 N(\dsp\to\taunu)=2217\pm83,\\
\end{equation}
where the error is statistical only. As a check, we fit the $E_{\rm ECL}$ distributions while floating the yield of each of the
three tau decay modes. The resulting yields are
\begin{eqnarray}
 N(\dsp\to\tauenu)&=&952\pm59,\nonumber\\
 N(\dsp\to\taumunu)&=&758\pm48,\\
 N(\dsp\to\taupinu)&=&496\pm35,\nonumber
\end{eqnarray}
where the errors are statistical only. The sum of individual yields is in good agreement with the total signal yield determined
in the simultaneous fit.

\subsection{\boldmath $\dsp\to e^+\nu_{e}$}

\begin{figure}[t!]
 \centering
 \includegraphics[width=1\textwidth]{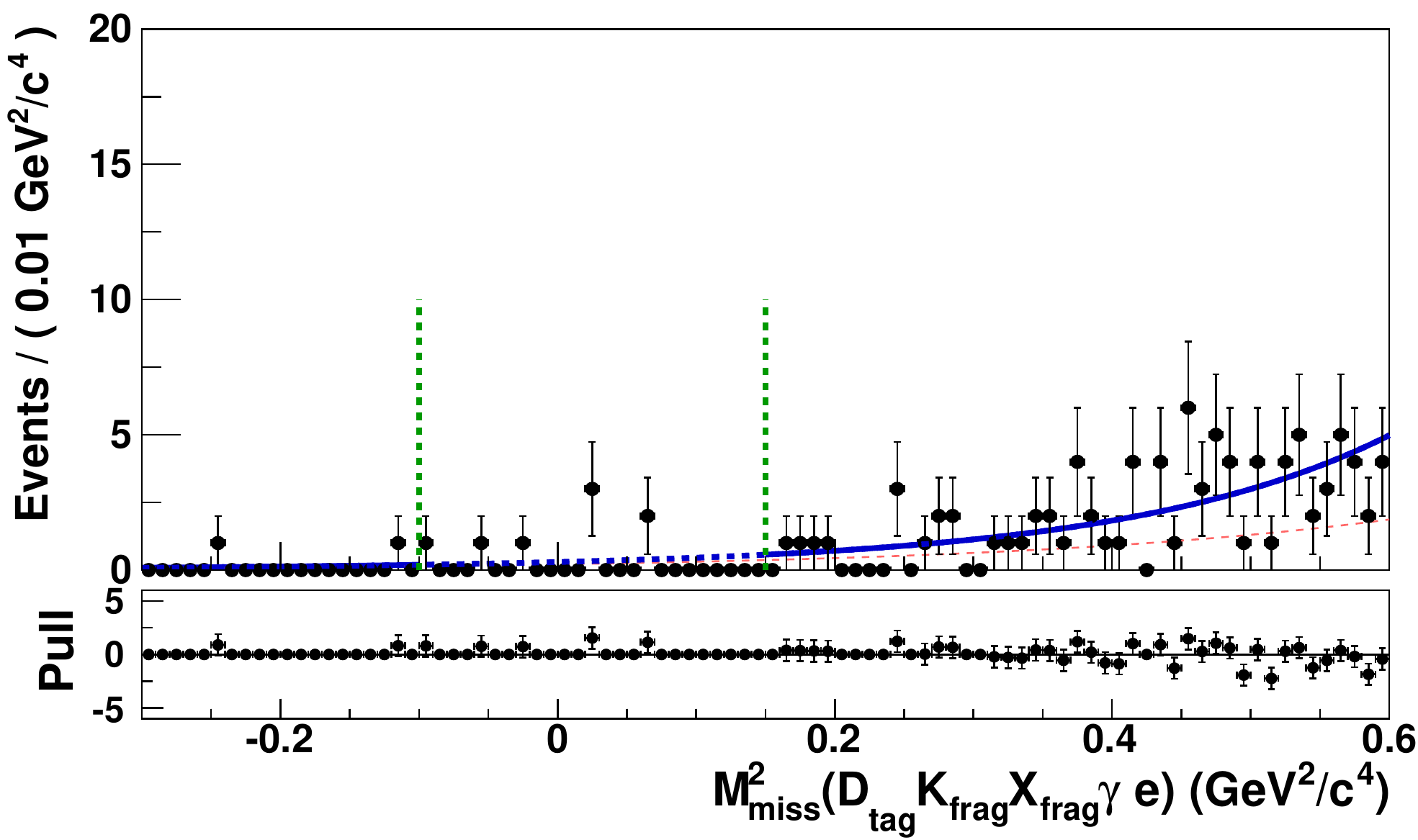}\\
\caption{The $\mmiss^2(\dtagkx\gamma e)$ distribution with superimposed fit results (blue lines). Events populating signal region in 
$\mmiss^2(\dtagkx\gamma e)$, denoted with two vertical dashed green lines, are excluded from the fit. The contribution of combinatorial
background candidates is indicated with the red dashed line. }
\label{fig:finalfit:enu:mc}
\end{figure}

We reconstruct $\ds^+\to e^+\nu_{e}$ decays in the same way as $\ds\to\tauenu$ decays, except that we focus on candidates 
populating the $\mmiss^2(\dtagkx\gamma e)$ region around zero.\footnote{The final states differ only in the number of neutrinos.}
To improve the purity, we require that $\eecl<0.5$~\gev and $p_{\rm miss}(\dtagkx\gamma e)>0.7$~\gevc in the laboratory frame. These two
requirements reject around 80\% of the background candidates while keeping around 75\% of the signal candidates in
the search window defined as $-0.10~\gevsqcq<\mmiss^2(\dtagkx\gamma e)<0.15$~\gevsqcq. The background in this window
is estimated by interpolating the observed yield from missing-mass-squared sidebands:
$-0.30~\gevsqcq<\mmiss^2(\dtagkx\gamma e)<-0.10$ \gevsqcq and $0.15~\gevsqcq<\mmiss^2(\dtagkx\gamma e)<0.60$ \gevsqcq. This is done
by performing a binned maximum likelihood fit in these sidebands, using a model consisting of a sum of two exponential
functions describing contributions of $\dsp\to\tauenu$ decays and combinatorial background. The shape and
normalization of the former are fixed to MC-based expectations. Contributions of other $\dsp$ decays are found to be negligible.
The shape parameter of the combinatorial background is fixed to
the value determined in a binned maximum likelihood fit to the $\mmiss^2(\dtagkx\gamma e)$ distribution of events from\linebreak $\mmiss(\dtagkx\gamma)$
sidebands. The normalization of the combinatorial background is a free parameter of the fit. Figure \ref{fig:finalfit:enu:mc}
shows the $\mmiss^2(\dtagkx\gamma e)$ distribution with superimposed fit. We observe
\begin{equation}
 n^{e\nu}_{\rm obs} = 8
 \label{eq:enu:nobs}
\end{equation}
in the $\mmiss^2(\dtagkx\gamma e)$ signal region (denoted by the two vertical lines in figure \ref{fig:finalfit:enu:mc}), which is in
good agreement with the estimated background 
\begin{equation}
 b^{e\nu}=8.7\pm0.9({\rm stat.})\pm0.8({\rm syst.}),
 \label{eq:enu:bkg}
\end{equation}
where the first error is statistical and the second is systematic. The latter is estimated by varying the yield
of $\dsp\to\tauenu$ decays and the combinatorial background parameters, changing both shape and normalization,
and repeating the fits.

\section{Determination of absolute branching fractions}
\label{sec:absbr}
For a given final state $f$, the absolute branching fraction of the $\dsp\to f$ decay is given by
\begin{equation}
 \br(\dsp\to f) = \frac{N(\dsp\to f)}{N^{\rm inc}_{\ds}\cdot f_{\rm bias}\cdot \varepsilon(\dsp\to f|{\rm incl.}~\dsp)}.
 \label{eq:abs_br_ideal}
\end{equation}
Here, $N^{\rm inc}_{\ds}$ is the number of inclusively reconstructed $\dsp$ mesons (see section \ref{sec:dsinclusive}), 
$N(\dsp\to f)$ is the number of reconstructed $\dsp\to f$ decays within the inclusive $\dsp$ sample 
(see section \ref{sec:exclusiverec}), and $\varepsilon(\dsp\to f|{\rm inc.}~\dsp)$ is the efficiency 
of reconstructing a $\dsp\to f$ decay within the inclusive $\dsp$ sample. 
MC studies show that the $\dsp$ inclusive reconstruction efficiency depends on the $\dsp$ decay mode 
and therefore the inclusively reconstructed $\dsp$ sample does not represent a truly inclusive sample 
of $\dsp$ mesons. 
To take this effect into account, a ratio of $\dsp$ inclusive reconstruction efficiency for $\dsp\to f$ decays,
$\varepsilon^{\rm inc}_{\ds\to f}$, and the average $\dsp$ inclusive reconstruction efficiency, 
$\overline{\varepsilon}{}_{\ds}^{\rm inc}=\sum_i {\cal B}(\dsp\to i)\varepsilon^{\rm inc}_{\ds\to i}$, is included in the
denominator of eq.~(\ref{eq:abs_br_ideal}): $\fbias = {\varepsilon^{\rm inc}_{\ds\to f}}/{\overline{\varepsilon}{}_{\ds}^{\rm inc}}$. 
The ratio $\fbias$ is taken from the MC sample.

Measured absolute branching fractions of the $\dskkpi$, $\kzk$, $\eta\pi^+$, $\munu$, and $\taunu$ decays are summarized together with the
corresponding signal yields and tag bias corrected efficiencies in table~\ref{tab:results}.
\begin{table}[t!]
 \centering
 \begin{tabular}{lccc}\hline\hline
  $\ds^+$ decay mode	& Signal yield				& $f_{\rm bias}\cdot\varepsilon$ [\%]	& $\br$ [\%] \\
  \hline\hline
  $K^-K^+\pi^+$				& $4094\pm123$				& $85.8$				& $5.06 \pm 0.15 \pm 0.21$\\
  $\overline{K}{}^0K^+$			& $2018\pm75\phantom{0}$		& $72.5$				& $2.95 \pm 0.11 \pm 0.09$\\
  $\eta\pi^+$				& $\phantom{0}788\pm59\phantom{0}$	& $45.8$				& $1.82 \pm 0.14 \pm 0.07$\\\hline
  $\mu^+\nu_{\mu}$			& $\phantom{0}492\pm26\phantom{0}$	& $98.2$				& $0.531\pm0.028\pm0.020$\\\hline
  $\tau^+\nu_{\tau}$ ($e$ mode)		& $\phantom{0}952\pm59\phantom{0}$	& $18.8$				& $5.37\pm0.33{}^{+0.35}_{-0.31}$\\
  $\tau^+\nu_{\tau}$ ($\mu$ mode)	& $\phantom{0}758\pm48\phantom{0}$	& $13.7$				& $5.86\pm0.37{}^{+0.34}_{-0.59}$\\
  $\tau^+\nu_{\tau}$ ($\pi$ mode)	& $\phantom{0}496\pm35\phantom{0}$	& $\phantom{0}8.7$			& $6.04\pm0.43{}^{+0.46}_{-0.40}$\\\hline
  $\tau^+\nu_{\tau}$ (combined)		& $2217\pm83\phantom{0}$		& $41.2$				& $5.70\pm0.21{}^{+0.31}_{-0.30}$\\\hline\hline
 \end{tabular}
 \caption{Signal yields, tag bias corrected efficiencies and measured branching fractions for all five studied $\ds^+$ decay modes. 
 The first uncertainty is statistical and the second is systematic. In the case of $\dsp\to\tau^+\nu_{\tau}$ decays the efficiencies include
 the branching fractions of the $\tau^+$ decay modes.}
 \label{tab:results}
\end{table}

\section{Systematic uncertainties}
\label{sec:syst}

Systematic uncertainties in the calculation of the branching fractions arise due to imperfect knowledge of the 
size of the inclusive $\dsp$ sample, the reconstruction efficiencies and the modeling of signal and background 
contributions in the fits from which the signal yields of $\dsp\to f$ decays are extracted. The estimated 
systematic uncertainties are itemized in table~\ref{tab:sytematics} and described below. 
\begin{table}[t!]
 \centering
 \renewcommand{\arraystretch}{1.2}
 \begin{tabular}{lcccccc} \hline\hline
  Source        	& $K^-K^+\pi^+$ [\%]	& $\kzk$ [\%]	& $\eta\pi^+$ [\%]	& $\enu$ [\%]	& $\munu$ [\%]	& $\taunu$ [\%]  \\\hline\hline
  Normalization 	& $\pm2.1$		& $\pm2.1$	& $\pm2.1$		& $\pm 2.1$	& $\pm 2.1$	& $\pm 2.1$\\
  Tag bias 		& $\pm 1.4$		& $\pm 1.4$	& $\pm 1.4$		& $\pm 1.4$	& $\pm 1.4$	& $\pm 1.4$\\
  Tracking 		& $\pm 1.1$		& $\pm 0.4$	& $\pm 0.4$		& $\pm 0.4$	& $\pm 0.4$	& $\pm 0.4$\\
  Particle ID		& $\pm 2.6$ 		& $\pm 0.8$	& $\pm 1.1$		& $\pm 1.9$	& $\pm 2.0$	& $\pm 1.7$\\
  Efficiency		& $\pm 0.7$ 		& $\pm 0.7$	& $\pm 1.4$		& $\pm 4.3$	& $\pm 1.8$	& $\pm 0.8$\\
  Dalitz model		& $\pm 1.1$		& --		& --			& --		& --		& --\\
  Fit model		& $\pm 0.8$		& $\pm 0.8$	& $\pm 2.2$		& --		& $\pm 0.2$	& $^{+3.3}_{-2.9}$\\
  $\dsp$ background	& --			& $\pm 0.6$	& $\pm 0.7$		& --		& $\pm 0.8$	& $\pm 2.8$\\
  $\tau$ cross-feed	& --			& --		& --			& --		& --		& $\pm 0.9$\\
  ${\cal B}(\tau\to X)$	& --			& --		& --			& --		& --		& $\pm 0.2$\\\hline\hline
  Total syst. 		& $\pm 4.1$		& $\pm 2.9$	& $\pm 3.9$		& $\pm5.4$	& $\pm 3.8$	& $^{+5.4}_{-5.2}$\\ \hline\hline
\end{tabular} 
\caption{Summary of systematic uncertainties for the branching fraction measurements of $\dsp$ decays. The total systematic error is 
 calculated by summing the individual uncertainties in quadrature. }
\label{tab:sytematics}
\end{table}

\paragraph{Normalization} The systematic error related to the normalization is assigned to be $\pm2.1$\% 
(where statistical and systematic errors given in eq.~(\ref{eq:inclds:yield}) are combined in quadrature) 
and is common for all studied $\dsp\to f$ decays. 

\paragraph{Tag bias} Possible differences in relative rates of individual $\dsp$ decay modes between MC simulation and 
data that impact the $\fbias$ calculation are estimated by studying the distributions of the number of charged particles 
and $\pi^0$'s, $N_{\rm ch} + N_{\pi^0}$, produced in $\dsp$ decays. We obtain the $N_{\rm ch} + N_{\pi^0}$ data distribution 
in the following way: first we count the number of remaining charged tracks, $N^{\rm reco}_{\rm ch}$, and $\pi^0$ candidates, 
$N_{\pi^0}^{\rm reco}$, that are not associated to the $\dtagkx\gamma$ candidate; in a second step, we determine, the inclusive 
$\dsp$ yields in bins of the $N^{\rm reco}_{\rm ch} + N^{\rm reco}_{\pi^0}$ by performing fits to the $\mmiss(\dtagkx\gamma)$ 
distributions. The obtained distribution of $N^{\rm reco}_{\rm ch} + N^{\rm reco}_{\pi^0}$ is proportional to the true distribution 
of $N_{\rm ch}+N_{\pi^0}$, but with a considerable amount of convolution.\footnote{E.g., a $\dsp$ daughter particle might not be 
reconstructed or a fake charged track or $\pi^0$ candidate may be counted.} We obtain the $N_{\rm ch} + N_{\pi^0}$ distribution
from the $N^{\rm reco}_{\rm ch} + N^{\rm reco}_{\pi^0}$ distribution using the singular value decomposition 
algorithm~\cite{Hocker:1995kb}. Finally, we estimate the ratio between the inclusive $\dsp$ reconstruction efficiencies in the data 
and MC samples,  
${\overline{\varepsilon}{}_{\ds}^{\rm inc}|_{\rm DATA}}/{\overline{\varepsilon}{}_{\ds}^{\rm inc}|_{\rm MC}}=0.9768\pm0.0134$,
using the unfolded data and MC $N_{\rm ch} + N_{\pi^0}$ distributions and the MC-determined dependence of the inclusive $\dsp$ 
reconstruction efficiency on $N_{\rm ch} + N_{\pi^0}$. The ratio is found to be consistent with unity within the uncertainty. 
Nevertheless, we correct the measured branching fractions by this factor and assign the error on this ratio as a source of systematic 
uncertainty ($\pm1.4\%$) that is common for all studied $\dsp$ decay modes.

\paragraph{Tracking, particle ID, and efficiency} The systematic errors in the $\dsp\to f$ reconstruction efficiencies 
arise from several sources. First, we assign a 0.35\% error per reconstructed charged track in the final state due to 
the uncertainty on the efficiency of the charged-track reconstruction estimated using partially 
reconstructed $D^{\ast+}\to D^0(K^0_S\pi^+\pi^-)\pi^+$ decays. The $e^+ e^-  \to e^+ e^- \ell^+ \ell^-$ ($\ell =$ $e$ or $\mu$) 
and $D^{\ast+}\to D^0(K^-\pi^+)\pi^+$ samples are used to estimate the lepton, kaon and pion identification corrections and 
pion (kaon) to lepton misidentification probabilities. Finally, we include statistical uncertainties of the MC-determined 
efficiencies $\fbias\cdot\varepsilon(\dsp\to f|{\rm inc.}~\dsp)$ as a source of systematic error. Since the 
branching fractions are determined relative to the number of inclusively reconstructed $\dsp$ mesons, the systematic 
uncertainties in the reconstruction of the inclusive $\dsp$ cancel.

\paragraph{Dalitz model} In the case of $\dskkpi$ decays, the reconstruction efficiency varies weakly across the $K^-K^+\pi^-$ 
Dalitz distribution. In calculating $\br(\dskkpi)$, we use the Dalitz-plot-integrated MC efficiency. The decay amplitude in 
the MC is simply the incoherent sum of all known resonant two-body contributions ($\phi\pi^+$, $\overline{K}{}^{\ast 0}K^+$, $f_0(980)\pi^+$ 
and others). We use a dedicated MC in which the $\dskkpi$ decay dynamics are simulated according to results of the Dalitz plot analysis 
performed by the CLEO collaboration \cite{Mitchell:2009aa}. The ratio of efficiencies from both samples is found to be consistent 
with unity within the uncertainty ($\pm1.1\%$) which is conservatively assigned as a source of systematic error.

\paragraph{Fit model} The systematic error due to limited statistics of the MC sample used to construct the histogram PDFs is 
evaluated by varying repeatedly the contents of all bins of all histogram PDFs within their statistical uncertainties and refitting. 
The RMS of the distribution of fit results is assigned as a systematic uncertainty due to the fit model. To estimate the systematic 
error due to the possible signal $\eecl$ shape difference between MC and data, we use the ratio of data to MC for the $\eecl$ histograms
of the background-subtracted $\dskkpi$ and $\dskzk$ samples to modify the signal PDF and repeat the fit. In a similar way, we estimate 
the systematic error due to the possible $\eecl$ shape difference of combinatorial background in MC and data by using the ratio of 
$\eecl$ histograms of $\dsp\to\taunu$ candidates populating the $\mmiss(\dtagkx\gamma)$ sidebands and modifying accordingly 
the combinatorial $\eecl$ PDF. For $\dsmunu$, $\dskzk$, and $\dsetapi$ decays, we vary the values of all fixed parameters 
within their uncertainties (taking correlations into account) and assign the differences with respect to the nominal 
fits as the fit model systematic uncertainty. In the case of $\dskkpi$, the fits are repeated by changing 
$\sigma_{\rm cal}^{\rm excl}$ within its uncertainty to assess this source of systematic error.

\paragraph{\boldmath$\dsp$ background} In the fits, we fix the contributions of certain background $\dsp$ decays to the expectations 
based on their measured branching fractions. We estimate the systematic error related to this by changing these branching fractions 
by their experimental errors \cite{Beringer:1900zz}. In the case of $\dsp\to\tauenu$ and $\dsp\to\taumunu$ decays, the systematic
uncertainty from this source originates mainly from $\dsp\to \overline{K}{}^0\ell\nu$ decays; in the case of $\dsp\to\taupinu$, from 
$\dsetapi$ and $\dsp\to\rho K^+$ decays. 

\paragraph{\boldmath$\tau$ cross-feed} In the nominal fit to the $\eecl$ distribution of reconstructed $\dsp\to\taunu$ decays,
we fix the cross-feed contributions relative 
to the signal contributions. We vary the ratios of cross-feed candidates within their uncertainties, repeat the fits, and take 
the differences from the nominal fits as the systematic uncertainty.

\paragraph{\boldmath ${\cal B}(\tau\to X)$} We include the uncertainties of the branching fractions of $\tau^+$ decays to 
$e^+\nu_e\overline{\nu}{}_{\tau}$, $\mu^+\nu_{\mu}\overline{\nu}{}_{\tau}$ and $\pi^+\overline{\nu}{}_{\tau}$~\cite{Beringer:1900zz} as 
the systematic errors on $\br(\dsp\to\taunu)$.

\section{Results}
\label{sec:results}

\subsection{\bf Branching fractions}
\begin{figure}[t!]
 \centering
 \includegraphics[width=0.99\textwidth]{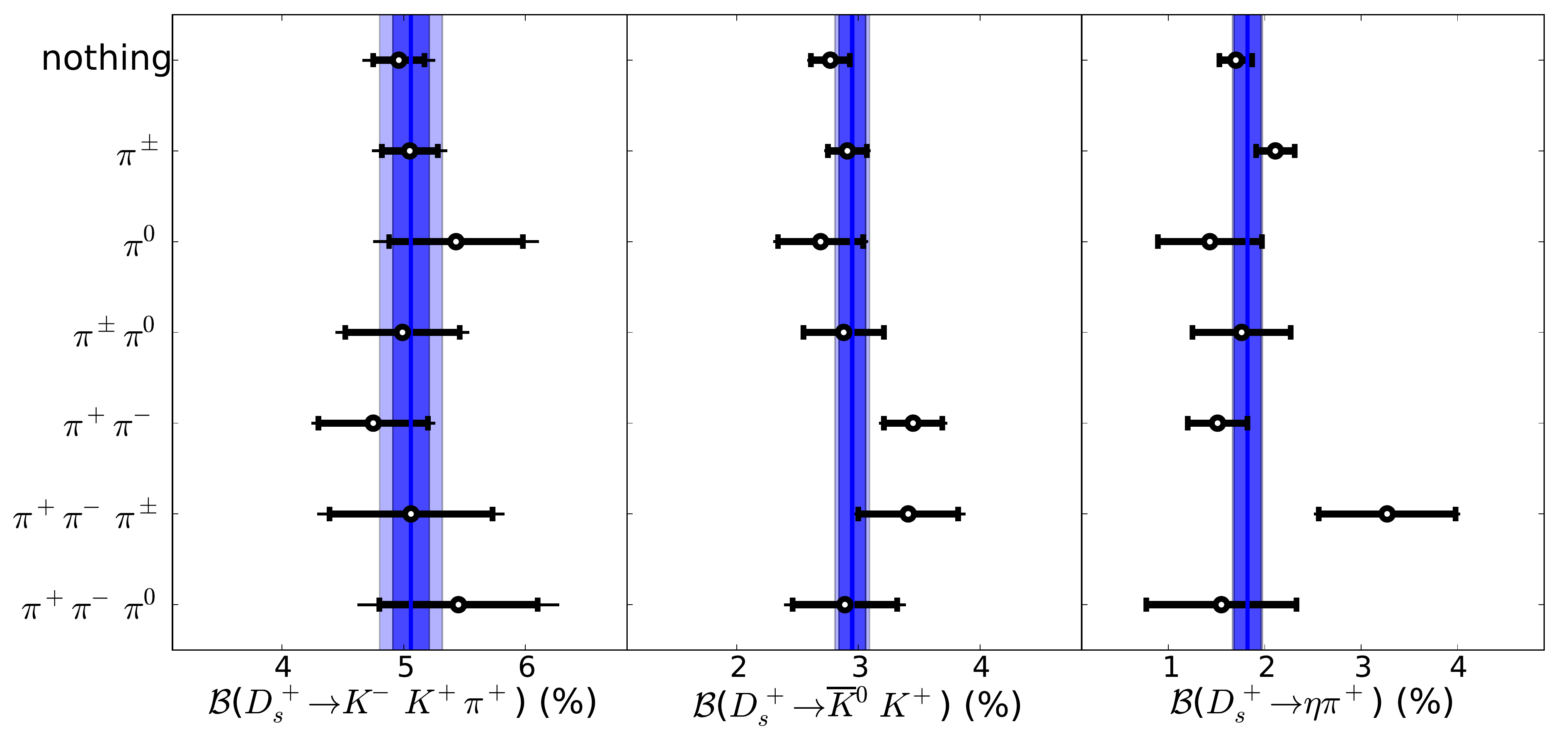}
 \caption{The measured $\br(\dskkpi)$ (left), $\br(\dskzk)$ (middle), and $\br(\dsetapi)$ (right) for each $\xfrag$ mode. 
 The measured branching fractions when all $\xfrag$ modes are combined are indicated with vertical blue lines with 
 statistical and total errors indicated as inner (dark blue) and outer (light blue) bands.}
 \label{fig:br:perxfrag}
\end{figure}

The signal yields of reconstructed $\dskkpi$, $\overline{K}{}^0K^+$, $\eta\pi^+$, $\mu^+\nu_{\mu}$, and $\tau^+\nu_{\tau}$ 
decays within the inclusive $\dsp$ sample are used to determine their absolute branching fractions (see also table~\ref{tab:results}). 
We measure the branching fractions of hadronic $\dsp$ decays to be
\begin{eqnarray}
 \br(\dskkpi)  & = & (5.06 \pm 0.15({\rm stat.}) \pm 0.21({\rm syst.}))\times 10^{-2},\\
 \br(\dskzk)   & = & (2.95 \pm 0.11({\rm stat.}) \pm 0.09({\rm syst.}))\times 10^{-2},\\
 \br(\dsetapi) & = & (1.82 \pm 0.14({\rm stat.}) \pm 0.07({\rm syst.}))\times 10^{-2}.
\end{eqnarray}
As a check, we determine the branching fractions of hadronic $\dsp$ decays for each $\xfrag$ mode separately. Figure~\ref{fig:br:perxfrag}
shows the measured branching fractions of the three hadronic decay modes, subdivided by $\xfrag$ mode. They are found to be in 
good agreement within uncertainties.

We measure the branching fractions of leptonic $\dsp$ decays to be
\begin{eqnarray}
 \br(\dsmunu)  & = & (5.31 \pm 0.28({\rm stat.}) \pm 0.20({\rm syst.}))\times 10^{-3},\\
 \br(\dsp\to\tau^+\nu_{\tau}) & = & (5.70 \pm 0.21({\rm stat.}) {}^{+0.31}_{-0.30}({\rm syst.}))\times 10^{-2},
\end{eqnarray}
where the first uncertainties are statistical and second systematic.  
The measured value of $\br(\dsmunu)$ is consistent with and represents a significant improvement over 
the previously measured value of $\br(\dsmunu)=(6.44 \pm 0.76({\rm stat.}) \pm 0.57({\rm syst.}))\times 10^{-3}$ by 
Belle~\cite{Widhalm:2007ws}. Table~\ref{tab:results} gives also 
the branching fractions of $\dsp\to\tau^+\nu_{\tau}$ decays as determined for individual $\tau^+$ decay modes. They are 
in good agreement within statistical uncertainties. As a test of lepton flavor universality, we determine 
the ratio of $\dsp$ leptonic decays to tau and muon to be
\begin{equation}
 R_{\tau/\mu}^{\ds} = 10.73\pm0.69({\rm stat.}){}^{+0.56}_{-0.53}({\rm syst.}),
\end{equation}
where we have taken into account that the common systematics between the two decay modes cancel in the ratio. The measured 
$R_{\tau/\mu}^{\ds}$ is in agreement with the SM value of $9.762\pm0.031$.

\begin{figure}[t!]
 \centering
 \includegraphics[width=0.7\textwidth]{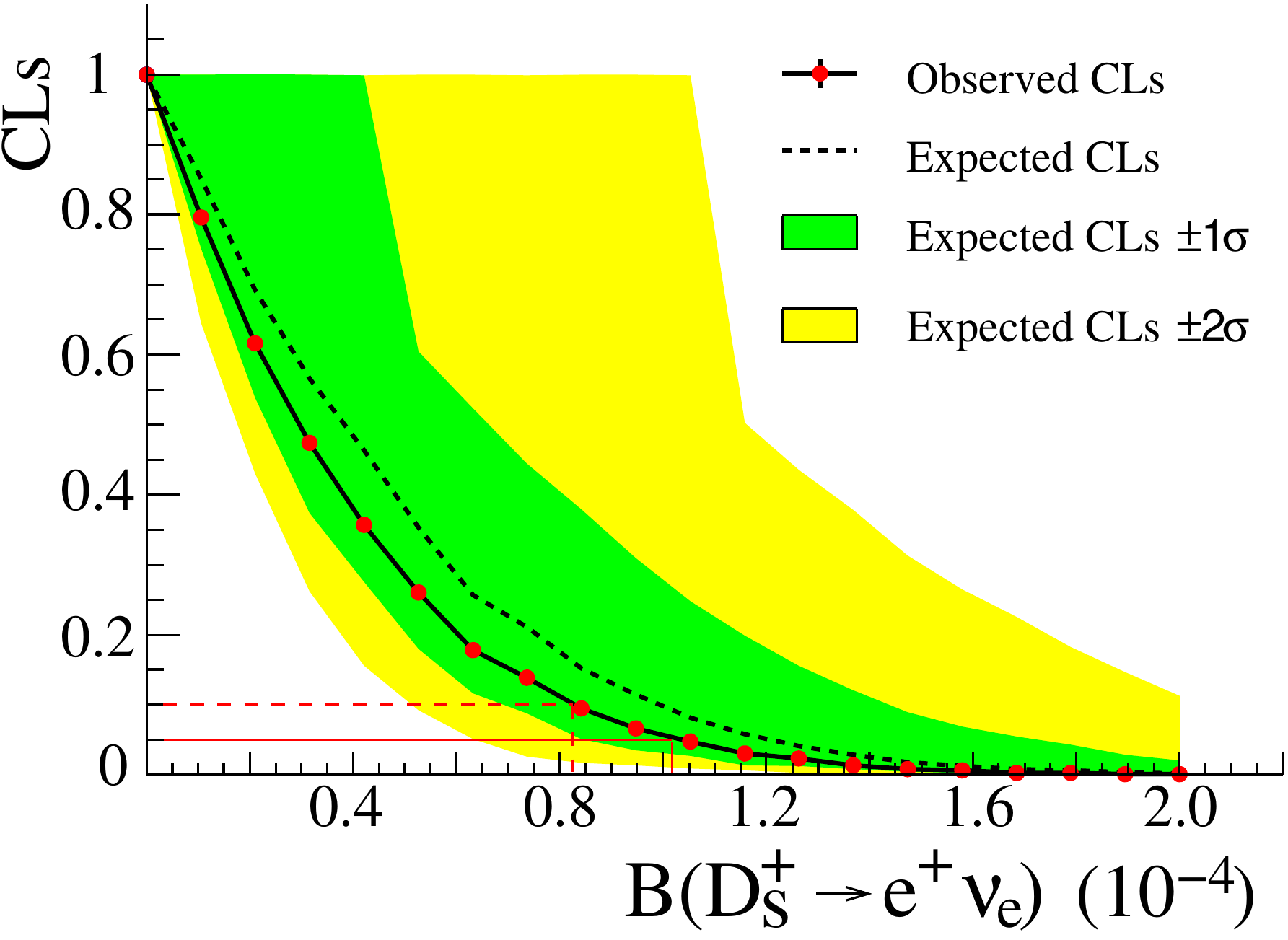}
 \caption{Observed (solid curve) and expected (dashed curve) \cls values as a function of $\br(\ds^+\to e^+\nu_{e})$. The green (yellow) 
 shaded area contains the $\pm1~\sigma$ ($\pm2~\sigma$) interval of possible results compatible with the expected value if only background 
 is observed. The upper limits at the 90\% (95\%) C.L. are indicated by the dashed (solid) line.}
 \label{fig:enu:cls}
\end{figure}

We find no evidence for $\dsp\to e^+\nu_{e}$ decays and set an upper limit on $\br(\dsp\to e^+\nu_{e})$ using the modified
frequentist approach (or \cls method)~\cite{Junk:1999kv,ALRead}. We construct test statistics from pseudo-experiments according 
to the signal plus background and background-only hypotheses~\cite{Moneta:2010pm}. For each pseudo-experiment, a 
likelihood ratio is
computed depending on the expected number of signal events for a given value of $\br(\dsp\to e^+\nu_{e})$ (according to 
eq.~(\ref{eq:abs_br_ideal})),
\begin{eqnarray}
 s^{e\nu} & = & N_{\rm incl.}^{\ds}\cdot\fbias\cdot\varepsilon(\dsp\to e^+\nu|{\rm incl.}~\dsp)\cdot\br(\ds^+\to e^+\nu_{e})\nonumber\\
   & = & (59370\pm3150({\rm stat.+syst.}))\cdot\br(\ds^+\to e^+\nu_{e}),
\end{eqnarray}
the expected number of background events, $b^{e\nu}$ (see eq.~(\ref{eq:enu:bkg})), and the observed number of events,
$n^{e\nu}_{\rm obs}$ (see eq.~(\ref{eq:enu:nobs})). The \cls is defined as the ratio of confidence levels, $\clsb/\clb$, where
the \clsb and \clb are the probabilities for signal-plus-background or background-only generated pseudo-experiments, respectively, 
to have a test-statistic value larger than or equal to that observed in the data. The 
observed distribution of $\cls$ as a function of branching fraction of $\ds^+\to e^+\nu_{e}$ decays is shown
in figure~\ref{fig:enu:cls}. The upper limits on $\br(\ds^+\to e^+\nu_{e})$ at 95 (90)\% confidence level (C.L.), which corresponds 
to $\cls = 0.05 ~(0.1)$, are
\begin{equation}
 \br(\ds^+\to e^+\nu_{e}) < 1.0~(0.83)\times10^{-4}~ \mbox{at} ~95~(90)\%~\mbox{C.L.}
\end{equation}
The systematic uncertainties of $s^{e\nu}$ and $b^{e\nu}$ are included in the \cls method using the techniques described in refs.~\cite{Junk:1999kv,ALRead}.
The sources of systematic uncertainties that are considered in the evaluation of $s^{e\nu}$ are listed in table \ref{tab:sytematics}.

\subsection{\bf\boldmath Extraction of the $\dsp$ meson decay constant}
The value of $\fds$ is determined from the measured branching fractions of leptonic $\dsp$ decays. 
Inverting eq.~(\ref{eq:brleptonic_sm}) yields
\[
\fds=\frac{1}{G_Fm_{\ell}\left( 1-\frac{m_{\ell}^2}{m_{\ds}^2}\right)\left|V_{cs}\right|} \sqrt{\frac{8\pi\br(\dsp\to\ell^+\nu_{\ell})}{m_{\ds}\tau_{\ds}}}.
\]
The external inputs necessary in the extraction of $\fds$ from the measured branching fractions are given in 
table~\ref{tab:fds:externalinput}. The CKM matrix element $|V_{cs}|$ is obtained from the well-measured 
$|V_{ud}|=0.97425\pm0.00022$ and $|V_{cb}|=(40.9\pm1.1)\times10^{-3}$ \cite{Beringer:1900zz} by using the relation 
$|V_{cs}|=|V_{ud}|-|V_{cb}|^2/2$, following the prescription given in the section {\it Decay constants of charged pseudoscalar mesons} in 
ref.~\cite{Beringer:1900zz}.
All but one of the external inputs are very precisely measured and do not introduce additional uncertainties; the exception is 
the $\dsp$ lifetime, $\tau_{\ds}$, which introduces an 0.70\% relative uncertainty.
\begin{table}[t]
 \centering
 \begin{tabular}{lr}\hline\hline
  Quantity & Value \\\hline\hline
  $m_{\ds}$  & 1.96849(32) GeV \\
  $m_{\tau}$ & 1.77682(16) GeV \\
  $m_{\mu}$  & 0.1056583715(35) GeV \\
  $\tau_{\ds}$ & 0.500(7) ps\\
  $G_F$ & $1.16637(1)\times 10^{-5}$ GeV$^{-2}$\\
  $|V_{cs}|$ & 0.97341(22)\\\hline\hline
 \end{tabular}
 \caption{Numerical values of external parameters used in extraction of $\fds$. All values are taken from ref.~\cite{Beringer:1900zz} except
 for $|V_{cs}|$ where we follow the prescription given in the section {\it Decay constants of charged pseudoscalar mesons} of ref.~\cite{Beringer:1900zz}.}
 \label{tab:fds:externalinput}
\end{table}
Table \ref{tab:fds:belle} summarizes the obtained values of $\fds$ using the $\ds^+\to\mu^+\nu_{\mu}$ and $\ds^+\to\tau^+\nu_{\tau}$ decays. The 
error-weighted average is 
\[
 \fds = (255.5\pm4.2({\rm stat.})\pm4.8({\rm syst.})\pm1.8(\tau_{\ds}))~\mbox{MeV},
\]
where the correlation of the systematic uncertainties between the $\mu^+\nu_{\mu}$ and $\tau^+\nu_{\tau}$ have been 
taken into account. This is the most precise measurement of $\fds$ to date. 
\begin{table}[t!]
 \centering
 \begin{tabular}{lc}
  $\dsp\to \ell^+\nu_{\ell}$  & $f_{D_s}$ [MeV] \\ \hline\hline
  $\munu$          & $249.8\pm6.6({\rm stat.})\pm4.7({\rm syst.})\pm1.7(\tau_{D_s})$ \\
  $\taunu$         & $261.9\pm4.9({\rm stat.})\pm7.0({\rm syst.})\pm1.8(\tau_{D_s})$ \\\hline
  Combination       & $255.5\pm4.2({\rm stat.})\pm4.8({\rm syst.})\pm1.8(\tau_{D_s})$ \\\hline\hline
 \end{tabular}
 \caption{Measured values of $\fds$ in $\munu$ and $\taunu$ decay modes and their combination.}
 \label{tab:fds:belle}
\end{table}

\section{Conclusions}
\label{sec:conclusion}
In conclusion, we measure the absolute branching fractions of hadronic
decays to $K^-K^+\pi^+$, $\kzk$, and $\eta\pi^+$, and of leptonic $\dsp$ decays to $\munu$ and $\taunu$ 
using the large data sample collected by the Belle experiment at the KEKB collider. 
The branching fractions of hadronic $\dsp$ decays are in agreement with the measurements performed by 
CLEO~\cite{Alexander:2008aa,Onyisi:2013pua}. The measurements of branching fractions of leptonic $\dsp$ decays are 
consistent with the previous measurements performed by the CLEO and BaBar 
collaborations~\cite{Alexander:2009ux,Naik:2009tk,Onyisi:2009th,delAmoSanchez:2010jg}. 
From the measured branching fractions of leptonic $\dsp$ decays, we determine the $\dsp$ meson decay constant, 
$\fds=(255.5\pm4.2(\rm stat.)\pm4.8(\rm syst.)\pm1.8(\tau_{D_s}))$~\mev, which represents the single most precise measurement to date
and is found to be in agreement with the most precise lattice QCD calculation~\cite{Davies:2010ip}. We find no evidence for 
$\dsp\to\enu$ decays and we set the most stringent upper limit of $\br(\ds^+\to e^+\nu_{e})<1.0~(0.83)\times10^{-4}$ at 95 (90)\% C.L.

\acknowledgments
We thank the KEKB group for the excellent operation of the
accelerator; the KEK cryogenics group for the efficient
operation of the solenoid; and the KEK computer group,
the National Institute of Informatics, and the 
PNNL/EMSL computing group for valuable computing
and SINET4 network support.  We acknowledge support from
the Ministry of Education, Culture, Sports, Science, and
Technology (MEXT) of Japan, the Japan Society for the 
Promotion of Science (JSPS), and the Tau-Lepton Physics 
Research Center of Nagoya University; 
the Australian Research Council and the Australian 
Department of Industry, Innovation, Science and Research;
Austrian Science Fund under Grant No. P 22742-N16;
the National Natural Science Foundation of China under
contract No.~10575109, 10775142, 10875115 and 10825524; 
the Ministry of Education, Youth and Sports of the Czech 
Republic under contract No.~MSM0021620859;
the Carl Zeiss Foundation, the Deutsche Forschungsgemeinschaft
and the VolkswagenStiftung;
the Department of Science and Technology of India; 
the Istituto Nazionale di Fisica Nucleare of Italy; 
The BK21 and WCU program of the Ministry Education Science and
Technology, National Research Foundation of Korea Grant No.\ 
2010-0021174, 2011-0029457, 2012-0008143, 2012R1A1A2008330,
BRL program under NRF Grant No. KRF-2011-0020333,
and GSDC of the Korea Institute of Science and Technology Information;
the Polish Ministry of Science and Higher Education and 
the National Science Center;
the Ministry of Education and Science of the Russian
Federation and the Russian Federal Agency for Atomic Energy;
the Slovenian Research Agency;
the Basque Foundation for Science (IKERBASQUE) and the UPV/EHU under 
program UFI 11/55;
the Swiss National Science Foundation; the National Science Council
and the Ministry of Education of Taiwan; and the U.S.\
Department of Energy and the National Science Foundation.
This work is supported by a Grant-in-Aid from MEXT for 
Science Research in a Priority Area (``New Development of 
Flavor Physics''), and from JSPS for Creative Scientific 
Research (``Evolution of Tau-lepton Physics'').

\bibliographystyle{JHEP}
\bibliography{PD1244_JHEP_preprint}{}

\providecommand{\href}[2]{#2}\begingroup\raggedright\begin{thebibliography}{10}

\bibitem{Brodzicka:2012jm}
{\bf Belle} Collaboration, J.~Brodzicka et~al., {\it {Physics Achievements from
  the Belle Experiment}},  {\em Prog.Theor.Exp.Phys.} {\bf 2012} (2012) 04D001,
  [\href{http://xxx.lanl.gov/abs/1212.5342}{{\tt arXiv:1212.5342}}].

\bibitem{Beringer:1900zz}
{\bf Particle Data Group} Collaboration, J.~Beringer et~al., {\it {Review of
  Particle Physics (RPP)}},  {\em Phys.Rev.} {\bf D86} (2012) 010001.

\bibitem{Adachi:2012mm}
{\bf Belle} Collaboration, I.~Adachi et~al., {\it {Measurement of $B^- \to
  \tau^- \bar{\nu}_\tau$ with a Hadronic Tagging Method Using the Full Data
  Sample of Belle}},  {\em Phys.Rev.Lett.} {\bf 110} (2013) 131801,
  [\href{http://xxx.lanl.gov/abs/1208.4678}{{\tt arXiv:1208.4678}}].

\bibitem{Hara:2010dk}
{\bf Belle} Collaboration, K.~Hara et~al., {\it {Evidence for $B^- \to \tau^-
  \bar{\nu}$ with a Semileptonic Tagging Method}},  {\em Phys.Rev.} {\bf D82}
  (2010) 071101, [\href{http://xxx.lanl.gov/abs/1006.4201}{{\tt
  arXiv:1006.4201}}].

\bibitem{Aaij:2012nna}
{\bf LHCb} Collaboration, R.~Aaij et~al., {\it {First Evidence for the Decay
  $B^0_s \to \mu^+\mu^-$}},  {\em Phys.Rev.Lett.} {\bf 110} (2013) 021801,
  [\href{http://xxx.lanl.gov/abs/1211.2674}{{\tt arXiv:1211.2674}}].

\bibitem{Akeroyd:2007eh}
A.~Akeroyd and C.~Chen, {\it {Effect of $H^{\pm}$ on $B^{\pm}\to
  \tau^{\pm}\nu_{\tau}$ and $D_{s}^{\pm}\to\mu^{\pm}\nu_{\mu}$,
  $\tau^{\pm}\nu_{\tau}$}},  {\em Phys.Rev.} {\bf D75} (2007) 075004,
  [\href{http://xxx.lanl.gov/abs/hep-ph/0701078}{{\tt hep-ph/0701078}}].

\bibitem{Akeroyd:2009tn}
A.~Akeroyd and F.~Mahmoudi, {\it {Constraints on charged Higgs bosons from
  $D_{s}^{\pm}\to\mu^{\pm}\nu$ and $D_{s}^{\pm}\to\tau^{\pm}\nu$}},  {\em JHEP}
  {\bf 0904} (2009) 121, [\href{http://xxx.lanl.gov/abs/0902.2393}{{\tt
  arXiv:0902.2393}}].

\bibitem{Barranco:2013tba}
J.~Barranco, D.~Delepine, V.~G. Macias, and L.~Lopez-Lozano, {\it {Constraining
  New Physics with $D$ meson decays}},
  \href{http://xxx.lanl.gov/abs/1303.3896}{{\tt arXiv:1303.3896}}.

\bibitem{Crivellin:2013wna}
A.~Crivellin, A.~Kokulu, and C.~Greub, {\it {Flavor-phenomenology of
  two-Higgs-doublet models with generic Yukawa structure}},  {\em Phys.Rev.}
  {\bf D87} (2013) 094031, [\href{http://xxx.lanl.gov/abs/1303.5877}{{\tt
  arXiv:1303.5877}}].

\bibitem{Filipuzzi:2012mg}
A.~Filipuzzi, J.~Portoles, and M.~Gonzalez-Alonso, {\it {U(2)$^5$ flavor
  symmetry and lepton universality violation in $W \to \tau \nu_{\tau}$}},
  {\em Phys.Rev.} {\bf D85} (2012) 116010,
  [\href{http://xxx.lanl.gov/abs/1203.2092}{{\tt arXiv:1203.2092}}].

\bibitem{Alexander:2009ux}
{\bf CLEO} Collaboration, J.~Alexander et~al., {\it {Measurement of $B(D_s^+
  \to \ell^+ \nu)$ and the Decay Constant $f_{D_s}^+$ From 600 $/pb^{-1}$ of
  $e^\pm$ Annihilation Data Near 4170 MeV}},  {\em Phys.Rev.} {\bf D79} (2009)
  052001, [\href{http://xxx.lanl.gov/abs/0901.1216}{{\tt arXiv:0901.1216}}].

\bibitem{Naik:2009tk}
{\bf CLEO} Collaboration, P.~Naik et~al., {\it {Measurement of the Pseudoscalar
  Decay Constant $f_{D_s}$ Using $D_s^+ \to\tau^+\nu$, $\tau^+\to\rho^+
  \overline{\nu}$ Decays}},  {\em Phys.Rev.} {\bf D80} (2009) 112004,
  [\href{http://xxx.lanl.gov/abs/0910.3602}{{\tt arXiv:0910.3602}}].

\bibitem{Onyisi:2009th}
{\bf CLEO} Collaboration, P.~Onyisi et~al., {\it {Improved Measurement of
  Absolute Branching Fraction of $D_s^+ \to \tau^+ \nu_{\tau}$}},  {\em
  Phys.Rev.} {\bf D79} (2009) 052002,
  [\href{http://xxx.lanl.gov/abs/0901.1147}{{\tt arXiv:0901.1147}}].

\bibitem{Widhalm:2007ws}
{\bf Belle} Collaboration, L.~Widhalm et~al., {\it {Measurement of
  $B(D_s^+\to\mu^+\nu)$}},  {\em Phys.Rev.Lett.} {\bf 100} (2008) 241801,
  [\href{http://xxx.lanl.gov/abs/0709.1340}{{\tt arXiv:0709.1340}}].

\bibitem{delAmoSanchez:2010jg}
{\bf BaBar} Collaboration, P.~del Amo~Sanchez et~al., {\it {Measurement of the
  Absolute Branching Fractions for $D^-_s\!\rightarrow\!\ell^-\bar{\nu}_{\ell}$
  and Extraction of the Decay Constant $f_{D_s}$}},  {\em Phys.Rev.} {\bf D82}
  (2010) 091103, [\href{http://xxx.lanl.gov/abs/1008.4080}{{\tt
  arXiv:1008.4080}}].

\bibitem{Davies:2010ip}
C.~Davies, C.~McNeile, E.~Follana, G.~Lepage, H.~Na, et~al., {\it {Update:
  Precision $D_s$ decay constant from full lattice QCD using very fine
  lattices}},  {\em Phys.Rev.} {\bf D82} (2010) 114504,
  [\href{http://xxx.lanl.gov/abs/1008.4018}{{\tt arXiv:1008.4018}}].

\bibitem{Bazavov:2011aa}
{\bf Fermilab Lattice Collaboration, MILC} Collaboration, A.~Bazavov et~al.,
  {\it {$B$- and $D$-meson decay constants from three-flavor lattice QCD}},
  {\em Phys.Rev.} {\bf D85} (2012) 114506,
  [\href{http://xxx.lanl.gov/abs/1112.3051}{{\tt arXiv:1112.3051}}].

\bibitem{Becirevic:2013mp}
D.~Becirevic, B.~Blossier, A.~Gerardin, A.~Le~Yaouanc, and F.~Sanfilippo, {\it
  {On the significance of B-decays to radially excited D}},  {\em Nucl.Phys.}
  {\bf B872} (2013) 313--332, [\href{http://xxx.lanl.gov/abs/1301.7336}{{\tt
  arXiv:1301.7336}}].

\bibitem{Blossier:2009bx}
{\bf ETM} Collaboration, B.~Blossier et~al., {\it {Pseudoscalar decay constants
  of kaon and $D$-mesons from $N_f=2$ twisted mass Lattice QCD}},  {\em JHEP}
  {\bf 0907} (2009) 043, [\href{http://xxx.lanl.gov/abs/0904.0954}{{\tt
  arXiv:0904.0954}}].

\bibitem{Bordes:2005wi}
J.~Bordes, J.~Penarrocha, and K.~Schilcher, {\it {$D$ and $D_s$ decay constants
  from QCD duality at three loops}},  {\em JHEP} {\bf 0511} (2005) 014,
  [\href{http://xxx.lanl.gov/abs/hep-ph/0507241}{{\tt hep-ph/0507241}}].

\bibitem{Lucha:2011zp}
W.~Lucha, D.~Melikhov, and S.~Simula, {\it {OPE, charm-quark mass, and decay
  constants of $D$ and $D_s$ mesons from QCD sum rules}},  {\em Phys.Lett.}
  {\bf B701} (2011) 82--88, [\href{http://xxx.lanl.gov/abs/1101.5986}{{\tt
  arXiv:1101.5986}}].

\bibitem{Badalian:2007km}
A.~Badalian, B.~Bakker, and Y.~Simonov, {\it {Decay constants of the
  heavy-light mesons from the field correlator method}},  {\em Phys.Rev.} {\bf
  D75} (2007) 116001, [\href{http://xxx.lanl.gov/abs/hep-ph/0702157}{{\tt
  hep-ph/0702157}}].

\bibitem{Hwang:2009qz}
C.-W. Hwang, {\it {SU(3) symmetry breaking in decay constants and
  electromagnetic properties of pseudoscalar heavy mesons}},  {\em Phys.Rev.}
  {\bf D81} (2010) 054022, [\href{http://xxx.lanl.gov/abs/0910.0145}{{\tt
  arXiv:0910.0145}}].

\bibitem{Fleischer:2010ay}
R.~Fleischer, N.~Serra, and N.~Tuning, {\it {A New Strategy for $B_s$ Branching
  Ratio Measurements and the Search for New Physics in $B^0_s \to \mu^+
  \mu^-$}},  {\em Phys.Rev.} {\bf D82} (2010) 034038,
  [\href{http://xxx.lanl.gov/abs/1004.3982}{{\tt arXiv:1004.3982}}].

\bibitem{Aaij:2011jp}
{\bf LHCb} Collaboration, R.~Aaij et~al., {\it {Measurement of $b$-hadron
  production fractions in $7~\rm{TeV}$ $pp$ collisions}},  {\em Phys.Rev.} {\bf
  D85} (2012) 032008, [\href{http://xxx.lanl.gov/abs/1111.2357}{{\tt
  arXiv:1111.2357}}].

\bibitem{Aaij:2013qqa}
{\bf LHCb} Collaboration, R.~Aaij et~al., {\it {Measurement of the
  fragmentation fraction ratio $f_{s}/f_{d}$ and its dependence on $B$ meson
  kinematics}},  {\em JHEP} {\bf 1304} (2013) 001,
  [\href{http://xxx.lanl.gov/abs/1301.5286}{{\tt arXiv:1301.5286}}].

\bibitem{Alexander:2008aa}
{\bf CLEO} Collaboration, J.~Alexander et~al., {\it {Absolute Measurement of
  Hadronic Branching Fractions of the $D_s^+$ Meson}},  {\em Phys.Rev.Lett.}
  {\bf 100} (2008) 161804, [\href{http://xxx.lanl.gov/abs/0801.0680}{{\tt
  arXiv:0801.0680}}].

\bibitem{Onyisi:2013pua}
{\bf CLEO} Collaboration, P.~Onyisi et~al., {\it {Improved Measurement of
  Absolute Hadronic Branching Fractions of the $D_s^+$ Meson}},
  \href{http://xxx.lanl.gov/abs/1306.5363}{{\tt arXiv:1306.5363}}.

\bibitem{Kurokawa:2001nw}
S.~Kurokawa and E.~Kikutani, {\it {Overview of the KEKB accelerators}},  {\em
  Nucl.Instrum.Meth.} {\bf A499} (2003) 1--7.

\bibitem{AbeKEKB}
T.~Abe et~al., {\it {Achievements of KEKB}},  {\em Prog.Theor.Exp.Phys.} {\bf
  2013} (2013) 03A001.

\bibitem{Abashian:2000cg}
{\bf Belle} Collaboration, A.~Abashian et~al., {\it {The Belle Detector}},
  {\em Nucl.Instrum.Meth.} {\bf A479} (2002) 117--232.

\bibitem{Lange:2001uf}
D.~Lange, {\it {The EvtGen particle decay simulation package}},  {\em
  Nucl.Instrum.Meth.} {\bf A462} (2001) 152--155.

\bibitem{Sjostrand:1993yb}
T.~Sj{\"{o}}strand, {\it {High-energy physics event generation with PYTHIA 5.7
  and JETSET 7.4}},  {\em Comput.Phys.Commun.} {\bf 82} (1994) 74--90.

\bibitem{Brun:1987ma}
R.~Brun, F.~Bruyant, M.~Maire, A.~McPherson, and P.~Zanarini, {\it {GEANT3}},
  1987.

\bibitem{Barberio:1993qi}
E.~Barberio and Z.~Was, {\it {PHOTOS: A Universal Monte Carlo for QED radiative
  corrections. Version 2.0}},  {\em Comput.Phys.Commun.} {\bf 79} (1994)
  291--308.

\bibitem{Widhalm:2006wz}
{\bf Belle} Collaboration, L.~Widhalm et~al., {\it {Measurement of $D^0 \to \pi
  \ell \nu$ ($K\ell\nu$) Form Factors and Absolute Branching Fractions}},  {\em
  Phys.Rev.Lett.} {\bf 97} (2006) 061804,
  [\href{http://xxx.lanl.gov/abs/hep-ex/0604049}{{\tt hep-ex/0604049}}].

\bibitem{Feindt:2006pm}
M.~Feindt and U.~Kerzel, {\it {The NeuroBayes neural network package}},  {\em
  Nucl.Instrum.Meth.} {\bf A559} (2006) 190--194.

\bibitem{Pivk:2004ty}
M.~Pivk and F.~R. Le~Diberder, {\it {\hbox{$_s$}${\cal P}$lot: A Statistical
  tool to unfold data distributions}},  {\em Nucl.Instrum.Meth.} {\bf A555}
  (2005) 356--369, [\href{http://xxx.lanl.gov/abs/physics/0402083}{{\tt
  physics/0402083}}].

\bibitem{Blobel}
V.~Blobel, ``{Smoothing of Poisson distributed data}.''
  http://www.desy.de/$\sim$blobel/splft.f.

\bibitem{Mendez:2009aa}
{\bf CLEO} Collaboration, H.~Mendez et~al., {\it {Measurements of $D$ Meson
  Decays to Two Pseudoscalar Mesons}},  {\em Phys.Rev.} {\bf D81} (2010)
  052013, [\href{http://xxx.lanl.gov/abs/0906.3198}{{\tt arXiv:0906.3198}}].

\bibitem{Hocker:1995kb}
A.~H{\"{o}}cker and V.~Kartvelishvili, {\it {SVD approach to data unfolding}},
  {\em Nucl.Instrum.Meth.} {\bf A372} (1996) 469--481,
  [\href{http://xxx.lanl.gov/abs/hep-ph/9509307}{{\tt hep-ph/9509307}}].

\bibitem{Mitchell:2009aa}
{\bf CLEO} Collaboration, R.~Mitchell et~al., {\it {Dalitz Plot Analysis of
  $D^+_s \to K^+K^-\pi^+$}},  {\em Phys.Rev.} {\bf D79} (2009) 072008,
  [\href{http://xxx.lanl.gov/abs/0903.1301}{{\tt arXiv:0903.1301}}].

\bibitem{Junk:1999kv}
T.~Junk, {\it {Confidence level computation for combining searches with small
  statistics}},  {\em Nucl.Instrum.Meth.} {\bf A434} (1999) 435--443,
  [\href{http://xxx.lanl.gov/abs/hep-ex/9902006}{{\tt hep-ex/9902006}}].

\bibitem{ALRead}
A.~L. Read, {\it {Presentation of search results: the CL$_{\rm s}$ technique}},
   {\em J.Phys.G:Nucl.Part.Phys.} {\bf 28} (2002) 2693--2704.

\bibitem{Moneta:2010pm}
L.~Moneta, K.~Belasco, K.~S. Cranmer, S.~Kreiss, A.~Lazzaro, et~al., {\it {The
  RooStats Project}},  {\em PoS} {\bf ACAT2010} (2010) 057,
  [\href{http://xxx.lanl.gov/abs/1009.1003}{{\tt arXiv:1009.1003}}].

\end{thebibliography}\endgroup
 
\end{document}